\documentclass[pdflatex,sn-nature]{sn-jnl}

\usepackage{lineno}
\usepackage{comment}
\usepackage{graphicx}%
\usepackage{multirow}%
\usepackage{amsmath,amssymb,amsfonts}%
\usepackage{amsthm}%
\usepackage{mathrsfs}%
\usepackage[title]{appendix}%
\usepackage[dvipsnames]{xcolor}
\usepackage{textcomp}%
\usepackage{manyfoot}%
\usepackage{booktabs}%
\usepackage{algorithm}%
\usepackage{algorithmicx}%
\usepackage{algpseudocode}%
\usepackage{listings}%
\usepackage{lmodern}
\usepackage{natbib}
\usepackage{anyfontsize}

\setcitestyle{super,open={},close={}}
\makeatletter
\newcommand\nonatspace{%
    \let\oriNAT@spacechar\NAT@spacechar\def\NAT@spacechar{\let\NAT@spacechar\oriNAT@spacechar}}
\makeatother
\usepackage[capitalise]{cleveref}
\usepackage{upgreek}
\usepackage[none]{hyphenat}
\usepackage{nth}
\usepackage{soul}

\newcommand{\Mtot}{{$\mathrm{M}_\mathrm{tot}$}}
\newcommand{\Mstar}{{{$\mathrm{M}_{\ast}$}}}

\newcommand{\tauV}{{$\hat{\tau}_{\scriptscriptstyle V}$}}

\newcommand{\logUs}{{$\log \mathrm{U}_{\scriptscriptstyle S}$}}
\newcommand{\Zstar}{{$\mathrm{Z}_{\ast}$}}
\newcommand{\Zmetal}{{$\mathrm{Z}$}}
\newcommand{\Zism}{{$\mathrm{Z}_\mathrm{ism}$}}
\newcommand{\age}{{$\mathrm{t}$}}

\newcommand{\fesc}{{$\mathrm{f}_\mathrm{esc }$}}

\newcommand{\Lya}{{Ly$\alpha$}}
\newcommand{\beagle}{{\tt BEAGLE}}

\begin{document}

\title[Article Title]{Spectroscopic confirmation of two luminous galaxies at a redshift of 14}

\author[1]{\fnm{Stefano} \sur{Carniani}}
\author[2]{\fnm{Kevin} \sur{Hainline}}
\author[3,4]{\fnm{Francesco} \sur{D'Eugenio}}
\author[5]{\fnm{Daniel J.} \sur{Eisenstein}}
\author[6,7]{\fnm{Peter} \sur{Jakobsen}}
\author[3,4]{\fnm{Joris} \sur{Witstok}}
\author[5]{\fnm{Benjamin D.} \sur{Johnson}}
\author[8]{\fnm{Jacopo} \sur{Chevallard}}
\author[3,4,9]{\fnm{Roberto} \sur{Maiolino}}
\author[2]{\fnm{Jakob M.} \sur{Helton}}
\author[10]{\fnm{Chris} \sur{Willott}}
\author[11]{\fnm{Brant} \sur{Robertson}}
\author[2]{\fnm{Stacey} \sur{Alberts}}
\author[12]{\fnm{Santiago} \sur{Arribas}}
\author[3,4]{\fnm{William M.} \sur{Baker}}
\author[13]{\fnm{Rachana} \sur{Bhatawdekar}}
\author[14,15]{\fnm{Kristan} \sur{Boyett}}
\author[8]{\fnm{Andrew J.} \sur{Bunker}}
\author[8]{\fnm{Alex J.} \sur{Cameron}}
\author[5]{\fnm{Phillip A.} \sur{Cargile}}
\author[16]{\fnm{Stéphane} \sur{Charlot}}
\author[17]{\fnm{Mirko} \sur{Curti}}
\author[18]{\fnm{Emma} \sur{Curtis-Lake}}
\author[2]{\fnm{Eiichi} \sur{Egami}}
\author[19]{\fnm{Giovanna} \sur{Giardino}}
\author[20]{\fnm{Kate} \sur{Isaak}}
\author[2]{\fnm{Zhiyuan} \sur{Ji}}
\author[8]{\fnm{Gareth C.} \sur{Jones}}
\author[21]{\fnm{Nimisha} \sur{Kumari}}
\author[22]{\fnm{Michael V.} \sur{Maseda}}
\author[1]{\fnm{Eleonora} \sur{Parlanti}}
\author[12]{\fnm{Pablo G.} \sur{Pérez-González}}
\author[23]{\fnm{Tim} \sur{Rawle}}
\author[2]{\fnm{George} \sur{Rieke}}
\author[2]{\fnm{Marcia} \sur{Rieke}}
\author[12]{\fnm{Bruno} \sur{Rodríguez Del Pino}}
\author[8,9]{\fnm{Aayush} \sur{Saxena}}
\author[3,4]{\fnm{Jan} \sur{Scholtz}}
\author[24]{\fnm{Renske} \sur{Smit}}
\author[2]{\fnm{Fengwu} \sur{Sun}}
\author[3,4]{\fnm{Sandro} \sur{Tacchella}}
\author[3,4]{\fnm{Hannah} \sur{\"Ubler}}
\author[1]{\fnm{Giacomo} \sur{Venturi}}
\author[25]{\fnm{Christina C.} \sur{Williams}}
\author[2]{\fnm{Christopher N. A.} \sur{Willmer}}
\affil[1]{\orgdiv{Scuola Normale Superiore, Piazza dei Cavalieri 7, I-56126 Pisa, Italy}}
\affil[2]{\orgdiv{Steward Observatory, University of Arizona, 933 N. Cherry Avenue, Tucson, AZ 85721, USA}}
\affil[3]{\orgdiv{Kavli Institute for Cosmology, University of Cambridge, Madingley Road, Cambridge, CB3 0HA, UK}}
\affil[4]{\orgdiv{Cavendish Laboratory, University of Cambridge, 19 JJ Thomson Avenue, Cambridge, CB3 0HE, UK}}
\affil[5]{\orgdiv{Center for Astrophysics $|$ Harvard \& Smithsonian, 60 Garden St., Cambridge MA 02138 USA}}
\affil[6]{\orgdiv{Cosmic Dawn Center (DAWN), Copenhagen, Denmark}}
\affil[7]{\orgdiv{Niels Bohr Institute, University of Copenhagen, Jagtvej 128, DK-2200, Copenhagen, Denmark}}
\affil[8]{\orgdiv{Department of Physics, University of Oxford, Denys Wilkinson Building, Keble Road, Oxford OX1 3RH, UK}}
\affil[9]{\orgdiv{Department of Physics and Astronomy, University College London, Gower Street, London WC1E 6BT, UK}}
\affil[10]{\orgdiv{NRC Herzberg, 5071 West Saanich Rd, Victoria, BC V9E 2E7, Canada}}
\affil[11]{\orgdiv{Department of Astronomy and Astrophysics University of California, Santa Cruz, 1156 High Street, Santa Cruz CA 96054, USA}}
\affil[12]{\orgdiv{Centro de Astrobiolog\'ia (CAB), CSIC–INTA, Cra. de Ajalvir Km.~4, 28850- Torrej\'on de Ardoz, Madrid, Spain}}
\affil[13]{\orgdiv{European Space Agency (ESA), European Space Astronomy Centre (ESAC), Camino Bajo del Castillo s/n, 28692 Villanueva de la Cañada, Madrid, Spain}}
\affil[14]{\orgdiv{School of Physics, University of Melbourne, Parkville 3010, VIC, Australia}}
\affil[15]{\orgdiv{ARC Centre of Excellence for All Sky Astrophysics in 3 Dimensions (ASTRO 3D), Australia}}
\affil[16]{\orgdiv{Sorbonne Universit\'e, CNRS, UMR 7095, Institut d'Astrophysique de Paris, 98 bis bd Arago, 75014 Paris, France}}
\affil[17]{\orgdiv{European Southern Observatory, Karl-Schwarzschild-Strasse 2, 85748 Garching, Germany}}
\affil[18]{\orgdiv{Centre for Astrophysics Research, Department of Physics, Astronomy and Mathematics, University of Hertfordshire, Hatfield AL10 9AB, UK}}
\affil[19]{\orgdiv{ATG Europe for the European Space Agency, ESTEC, Noordwijk, The Netherlands}}
\affil[20]{\orgdiv{European Space Agency, ESTEC, Noordwijk, Netherlands}}
\affil[21]{\orgdiv{AURA for European Space Agency, Space Telescope Science Institute, 3700 San Martin Drive. Baltimore, MD, 21210}}
\affil[22]{\orgdiv{Department of Astronomy, University of Wisconsin-Madison, 475 N. Charter St., Madison, WI 53706 USA}}
\affil[23]{\orgdiv{European Space Agency (ESA), European Space Astronomy Centre (ESAC), Camino Bajo del Castillo s/n, 28692 Villafranca del Castillo, Madrid, Spain}}
\affil[24]{\orgdiv{Astrophysics Research Institute, Liverpool John Moores University, 146 Brownlow Hill, Liverpool L3 5RF, UK}}
\affil[25]{\orgdiv{NSF’s National Optical-Infrared Astronomy Research Laboratory, 950 North Cherry Avenue, Tucson, AZ 85719, USA}}
\affil[]{}
\affil[]{}
\affil[]{}
\affil[]{}
\affil[]{}
\affil[]{}



\abstract{\unboldmath
The first observations of JWST have revolutionized our understanding of the Universe by identifying for the first time galaxies at $z\sim13$\cite{Curtis-Lake:2023, Wang:2023, Hainline:2024}. In addition, the discovery of many luminous galaxies at Cosmic Dawn ($z>10$) has suggested that galaxies developed rapidly, in apparent tension with many standard models \cite{Finkelstein:2023a, Harikane:2023, Casey:2023, Robertson:2023a, Donnan:2024}. However, most of these galaxies lack spectroscopic confirmation, so their distances and properties are uncertain. We present JADES JWST/NIRSpec spectroscopic confirmation of two luminous galaxies at redshifts of $z=14.32^{+0.08}_{-0.20}$ and $z=13.90\pm0.17$. The spectra reveal ultraviolet continua with prominent Lyman-$\alpha$ breaks but no detected emission lines.
This discovery proves that luminous galaxies were already in place 300~million years after the Big Bang and are more common than what was expected before JWST. The most distant of the two galaxies is unexpectedly luminous and is spatially resolved with a radius of 260 parsecs. 
Considering also the very steep ultraviolet slope of the second galaxy, we conclude that both are dominated by stellar continuum emission, showing that the excess of luminous galaxies in the early Universe cannot be entirely explained by accretion onto black holes. Galaxy formation models will need to address the existence of such large and luminous galaxies so early in cosmic history.}

\maketitle

JWST NIRSpec\cite{Jakobsen:2022} spectroscopic observations have  recently targeted three candidate galaxies at $z > 14$, selected within the JWST Advanced Deep Extragalactic Survey (JADES) campaigns\cite{Rieke:2023,Eisenstein:2023}.  
These galaxies were photometrically identified from within the 58 square arcminute observations of the GOODS-S field through JWST observations with up to 13 NIRCam and 7 MIRI filters \cite{Hainline:2023a, Williams:2023, Robertson:2023a}. 
Based on photometry from the Hubble Space Telescope (HST) and Cycle 1 JWST/NIRCam data, the probability of these galaxies being low-redshift interlopers was less than 1\% \cite{Hainline:2023a}.
By happenstance, the brightest of these three candidate galaxies (hereafter: JADES-GS-z14-0) is located at a projected distance of only $0.4$~arcsec from a foreground galaxy, and this interloper is at a redshift where its Balmer break is spectrally coincident with the observed photometric Lyman-$\alpha$ break of the distant galaxy.
For this reason, and due to its high inferred luminosity at the photometric redshift, JADES-GS-z14-0 was previously considered a low-redshift interloper with a peculiar spectral energy distribution \cite{Hainline:2023a, Williams:2023}. The ``low-redshift solution" was later disfavored from the analysis \cite{Robertson:2023a} of the JWST/NIRCam observations carried out in the JADES Origins Field program \cite{Eisenstein:2023}, which included additional deep medium-band NIRCam observations that substantially strengthened the case for the source being at high redshift.

The three galaxies were observed with NIRSpec in multi-object spectroscopic mode \cite{Ferruit:2022}, within a single NIRSpec field of view of 9 square arcmin, with both the low-resolution prism and all three medium-resolution gratings probing the wavelength range $0.6-5.2~\mathrm{\mu m}$  with spectral resolving powers ${\mathrm R \sim 100}$ and ${\mathrm R  \sim 1000}$, respectively. 
Owing to both the low luminosity of the source and NIRSpec slit losses, the faintest candidate is not significantly detected in the NIRSpec observations (see Methods), so hereon we focus on the other two galaxies, JADES-GS-z14-0 and JADES-GS-z14-1, which have been unambiguously detected in the prism spectra. 
 
\begin{figure}
    \centering
    \includegraphics[width=1\linewidth]{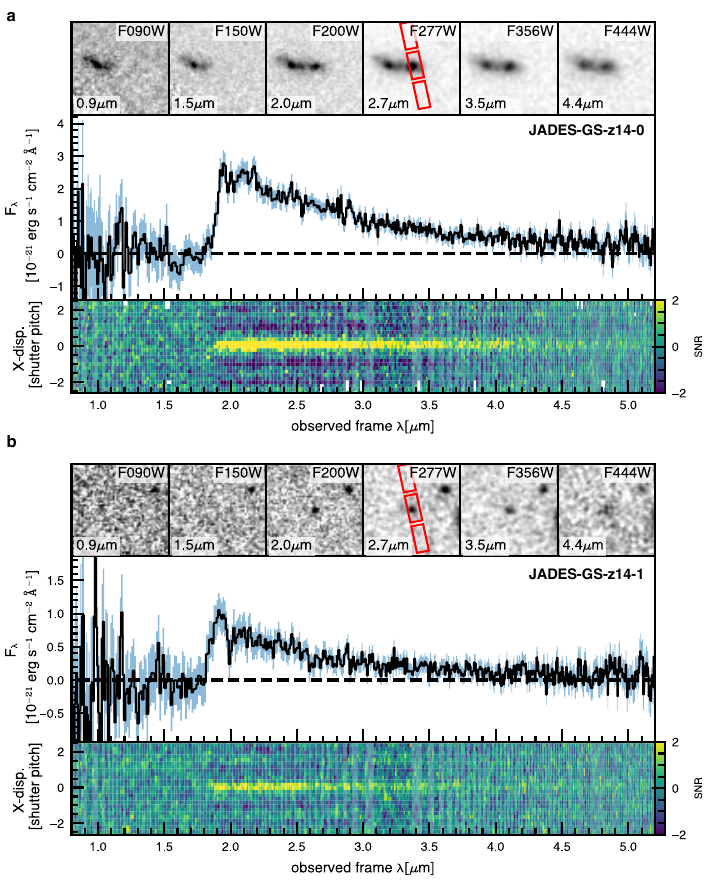}\\

    \caption{{\bf Spectra of the two the two $z\sim14$ galaxies. a-b,} NIRSpec prism (R=100) spectra for  JADES-GS-z14-0 ({\bf a}) and JADES-GS-z14-1 ({\bf b}). For each galaxy, the central panel shows the 1D spectrum (black) and the associated 1$\sigma$ uncertainty (light blue). The bottom panels display the 2D spectrum of the signal-to-noise ratio to better highlight the contrast across the break at $\sim1.8\,\mu$m. 
    Inset stamps in the top panels are cutouts of some of the NIRCam JADES images. The NIRSpec 3-shutter slitlets are shown in red in each F277W image.}
    \label{fig:spectra}
\end{figure}

Figure~\ref{fig:spectra} shows the prism spectra of JADES-GS-z14-0 and JADES-GS-z14-1; there are no prominent emission lines, but both galaxies display a clear break in the flux density, with no flux detected blueward of 1.85~$\mu$m; the sharpness of which  can only be explained as a Lyman-$\alpha$ break \cite{Curtis-Lake:2023, Wang:2023}, placing both galaxies at $z \sim 14$.

We have also obtained the spectrum of the low-redshift galaxy $0.4$~arcsec East of JADES-GS-z14-0, which has revealed multiple prominent emission lines (e.g. [OIII]$\lambda\lambda$4959,5007 and H$\alpha$) placing this projected nearby source at a redshift of $z = 3.475$ (see Methods). At this redshift, its Balmer break is at 1.62~$\mu$m, excluding the possibility that the sharp break in the flux density at 1.85~$\mu$m observed in the spectrum of JADES-GS-z14-0 is caused by contamination from the nearby foreground source.  The presence of the nearby low-redshift galaxy, however, mildly boosts the luminosity of JADES-GS-z14-0 via gravitational lensing. We have verified that the magnification factor is less than a factor of 1.2 (see Methods).

A redshift determination for galaxies within the Epoch of Reionization based solely on the Lyman-$\alpha$ break is sensitive to the absorption of neutral hydrogen along the line of sight \cite{Curtis-Lake:2023, DEugenio:2023, Hainline:2024}.
We have thus estimated the redshift of the two galaxies by parameterizing the rest-frame UV continuum emission with a power law of the form $F_{\lambda}\propto\lambda^{\beta}$ and taking into account the multiple physical processes that can shape the Lyman break profile in the prism spectra (see Methods).
The redshifts we have recovered from our best-fitting models are 
$z=14.32^{+0.08}_{-0.20}$ and $z=13.90\pm0.17$ for JADES-GS-z14-0 and JADES-GS-z14-1, respectively. 

In the redshift range inferred by fitting the Lyman break profile, we have also found a tentative detection of  CIII]$\lambda\lambda1907,1909$ (hereafter: CIII]) emission at 2.89~$\mu$m in  JADES-GS-z14-0 (see Methods) at a level of significance of 3.6$\sigma$. If confirmed in future NIRSpec observations, this line yields a redshift of $14.178\pm0.013$ and the presence of damped Lyman-$\alpha$ absorption (DLA) with a neutral hydrogen column density of $\log_{10}(N_\mathrm{HI}/\mathrm{cm^{-2}})=22.23\pm0.08 $ is necessary to match the wavelength and shape of the Lyman-$\alpha$ break. 

\begin{figure}
    \centering
    \includegraphics[width=0.5\linewidth]{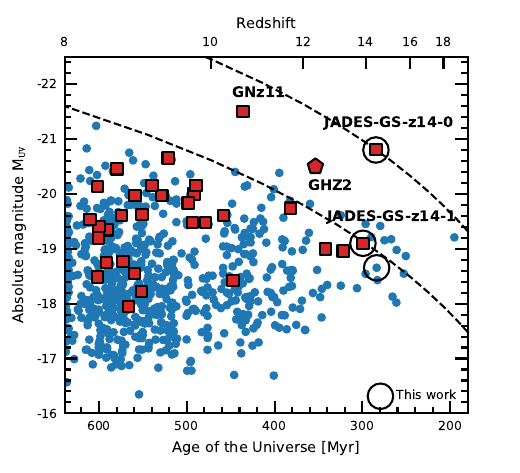}
\textbf{}    \caption{{\bf UV absolute magnitudes of galaxies at $z>8$}.  Blue circles are candidate high-$z$ galaxies in the GOODS-S and GOODS-N identified in JADES \cite{Hainline:2023a}, while red squares are the spectroscopically confirmed galaxies \cite{Bunker:2023a, DEugenio:2024}. For comparison, we also report the galaxy GHZ2 \cite{Castellano:2024, Zavala:2024} from the GLASS survey (red pentagon). Empty black circles highlight the targets analyzed in this work. The relatively low number of galaxies near $z=10$ is an artifact of photometric-redshift selections. Dashed lines illustrate a semi-empirical luminosity evolution ($\propto(1+z)^{4.5}$) of haloes of a given comoving abundances. }
    \label{fig:muv}
\end{figure}

These are the earliest galaxies with spectroscopically confirmed redshifts, exceeding the previous high marks of $z=13.2$ \cite{Curtis-Lake:2023, Hainline:2024} and $z=13.07$ \cite{Wang:2023}. Additionally, these two galaxies are luminous with a rest-frame UV absolute luminosity at 1500~\AA\  of $M_\mathrm{UV}=-20.81$ and $M_\mathrm{UV}=-19.00$, respectively. We particularly highlight JADES-GS-z14-0, which despite its redshift is the third most UV luminous of the 700 $z>8$ candidates in JADES, two times more luminous than GHZ2 \cite{Castellano:2024, Zavala:2024}, and only a factor of two less luminous than GN-z11 \cite{Oesch:2016, Bunker:2023}. We illustrate the distribution of UV luminosity and redshift in Fig.~\ref{fig:muv}.  We stress that the high luminosity is particularly important in view of the rapidly evolving halo mass function expected in cold-dark-matter cosmology. From a N-body simulation run with Abacus \cite{Maksimova:2021}, we estimate that the halo mass threshold required to yield a fixed comoving abundance varies as $(1+z)^{-6}$ for this region of mass and redshift.  A dimensional scaling for luminosity would be halo mass divided by the age of the Universe, which is scaling as $(1+z)^{-3/2}$, yielding a simple baseline that luminosities might scale as $(1+z)^{-4.5}$.  Overplotting such a scaling on Fig.~\ref{fig:muv} shows how remarkable JADES-GS-z14-0 is: it shows most dramatically that some astrophysical processes are creating a deviation from the dimensional scaling of halo mass and the Hubble time. Even JADES-GS-z14-1, while more similar in $M_\mathrm{UV}$ to the lower redshift family, is distinctively luminous by this metric.  We, therefore, argue that these two galaxies, and particularly JADES-GS-z14-0, provide a crisp spectroscopic confirmation to the trend that has been inferred several times from photometric samples \cite{Casey:2023, Finkelstein:2023a, Donnan:2024, Robertson:2023a} that the galaxy UV luminosity function evolves slowly, with more luminous galaxies at high redshift than predicted in a variety of pre-JWST predictions. Having established the remarkable redshifts and luminosities of these sources, we now turn to a more detailed analysis of them.

From the spectrum redward of the break, we measure a power-law index $\beta$, also known as the UV slope, of $-2.20\pm0.07$  and  $-2.71\pm{0.19}$ for JADES-GS-z14-0 and JADES-GS-z14-1, respectively. These results indicate that the emission is dominated by a relatively young ($<300$~Myr) stellar population and low dust attenuation \cite{Tacchella:2022, Cullen:2023, Topping:2023}. 
We note that the stellar UV slope could be also modified by two-photon and free-bound nebular continuum emission \cite{Robertson:2010, Bouwens:2010a, Raiter:2010, Tacchella:2022, Cullen:2023, Topping:2023, Cameron:2023}. However, we can rule out a strong two-photon contribution in our galaxies due to the lack of the characteristic peak at 1500~\AA\ \cite{Cameron:2023}. The absence of emission lines disfavors free-bound emission, but this possibility cannot be fully ruled out because, at $z\sim14$, NIRSpec does not cover the Balmer break nor any Balmer emission lines.

\begin{table}
    \centering
    \caption{Galaxy properties inferred from NIRSpec data corrected for slit-losses based on NIRCam fluxes.}
    \begin{tabular}{l c c}
    \hline
    \hline

ID & JADES-GS-z14-0 & JADES-GS-z14-1 \\
extended ID & JADES-GS-53.08294-27.85563 & JADES-GS-53.07427-27.88592  \\
NIRCam ID & 183348  & 18044 \\
RA[ICRS]  &  3:32:19.905   &  3:32:17.825  \\
DEC[ICRS]   & -27:51:20.27  &  -27:53:09.34 \\
\hline
redshift & $14.32^{+0.08}_{-0.20}$ &  $13.90\pm0.17$ \\
UV slope $\beta$ & $-2.20\pm0.07$  &  $-2.71\pm{0.19}$\\
$M_\mathrm{UV}$  & $-20.81\pm0.16$$^\dagger$  &  $-19.0\pm0.4$\\
UV radius ($r_\mathrm{UV}$) [pc] & $260\pm20$ & $<160$\\

$\log_{10}(M_\mathrm{star} / \textrm{M}_\odot)$$^\star$  & $8.6_{-0.2}^{+0.7}$$^\dagger$  & $8.0_{-0.3}^{+0.4}$  \\
$\text{SFR}_{100}~[\textrm{M}_\odot \,\textrm{yr}^{-1}$]& $4_{-3}^{+9}$$^\dagger$  & $1.2_{-0.9}^{+0.7}$ \\
$\text{SFR}_{10}~[\textrm{M}_\odot \,\textrm{yr}^{-1}$]& $19\pm6$$^\dagger$  & $2_{-0.4}^{+0.7}$ \\
$\text{sSFR}_{10}~[\textrm{Gyr}^{-1}$]& $45_{-35}^{+56}$  & $18_{-38}^{+75}$ \\
$A_\mathrm{V}$ [mag]&  $0.31_{-0.07}^{+0.14}$ & $0.20_{-0.07}^{+0.11}$  \\
$\log_{10}(Z/ Z_\odot)$ &  $-1.5_{-0.4}^{+0.7}$  &  $-1.1_{-0.5}^{+0.6}$  \\
$\text{f}_\text{esc}^\text{LyC}$ &  $0.84_{-0.16}^{+0.09}$  &  $0.63_{-0.29}^{+0.25}$  \\

    \hline
    \end{tabular}
{\bf Note}:$^\star$ uncertainties refer only to the internal statistical errors of our model. Stellar mass is sensitive to the assumptions on star-formation history\cite{Helton:2024}; $^\dagger$ corrected for gravitational lensing amplification of $\mu=1.17$ (see Methods).
    \label{tab:summary}
\end{table}

The physical properties of the two galaxies have been inferred via spectro-photometric modeling of their spectral energy distributions (SEDs) within a Bayesian framework. The details of the modeling and the posterior distribution of free parameters are discussed in the Methods section, while the galaxy properties are reported in Table~\ref{tab:summary}.
The inferred star-formation history indicates that these galaxies have grown their masses over the last 100 Myr, implying that the observed stellar population started forming at $z \sim 20$ with a rapid growth up to $z\sim14$ \cite{Helton:2024}. We also note that the SED modeling favors a high escape fraction of ionizing photons ($\text{f}_\text{esc}^\mathrm{
 LyC }>0.35$) to reproduce the blue UV slopes and the absence of emission lines in both galaxies.

\begin{figure}
    \centering
    \includegraphics[width=1\linewidth]{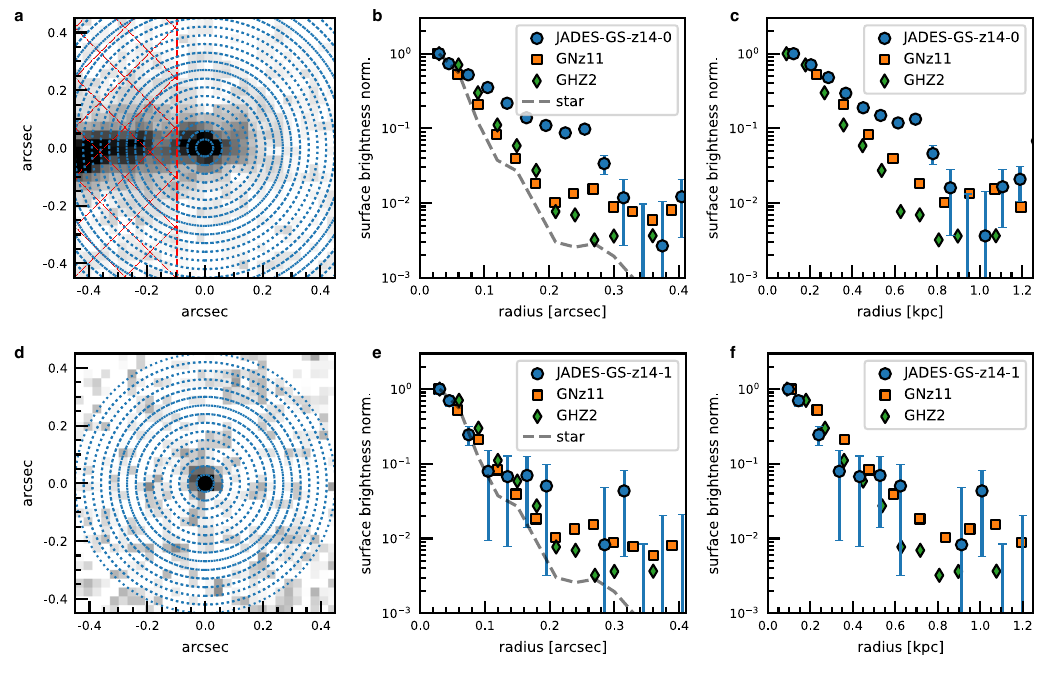}
    
    \caption{{\bf Galaxy size. a--f,} the normalized radial profile of the observed surface brightness at 2$\mu\mathrm{m}$ of JADES-GS-z14-0 ({\bf a,b,c}) and JADES-GS-z14-1({\bf d,e,f}). From left to right, the panels show the NIRCam image in the F200W filter ({\bf a,d}), the surface brightness profile in arcsec units ({\bf b,e}), and the surface brightness profile in kiloparsec units ({\bf c,f}). The circularized light profiles are extracted from the rings marked with blue concentric circles in the NIRCam image, and the dashed region in the JADES-GS-z14-0 image marks the part that was masked to remove the contamination of the neighboring $z=3.45$ galaxy. The comparison with the radial profile of a star in the field of view (gray dashed curve), GNz11 \cite{Tacchella:2023} (orange squares) and GHZ2 \cite{Castellano:2024} (green diamonds) shows that JADES-GS-z14-0 is significantly more extended than the most luminous galaxies previously known at $z>10$.}

    \label{fig:spatial_extension}
\end{figure}

The NIRCam images of JADES-GS-z14-0 clearly show that the source is extended, while JADES-GS-z14-1 is more compact. Fig.~\ref{fig:spatial_extension} shows the radial profile of the emission at 2~$\mu$m of the two galaxies. The radial surface brightness profile of JADES-GS-z14-0 exhibits emission extended up to 1 kpc, significantly beyond the point spread function of JWST. 
We also note that the profile is significantly more extended than the UV emission of the two more luminous galaxies at $z>10$: GN-z11 \cite{Bunker:2023, Maiolino:2023} and GHZ2 \cite{Castellano:2024, Zavala:2024}.
Using {\tt ForcePho} (see Methods) to fit the imaging, we find that the galaxy is well fit by an elliptical exponential profile with a deconvolved half-light radius ($r_\mathrm{UV}$) of $0.079\pm0.006$~arcsec and $260\pm20$ pc. This large size implies that the UV light of JADES-GS-z14-0 is produced mainly by a spatially extended stellar population, excluding a dominant contribution by an active galactic nucleus (AGN). This differs from other more compact high-luminosity galaxies, where some studies have suggested that an unobscured AGN is dominating the UV light \cite{Maiolino:2023, Harikane:2024}.

The rest-frame UV emission of JADES-GS-z14-1 appears compact and marginally resolved by the NIRCam point-spread function. 
The forward modeling of the light profile returns an upper limit on $r_\mathrm{UV}<160$~pc, which agrees with the compact size determined for other low-luminosity $z>10$ galaxies \cite{Robertson:2023, Robertson:2023a, Hainline:2024}. For this galaxy, the morphological analysis is not sufficient to exclude the presence of a luminous AGN, but the inferred UV slope of $-2.71\pm0.19$ suggests that the light is mainly emitted by stars in the galaxy.  The slope expected for the emission of an AGN accretion disk is, on average, of the order of --2.3 or shallower \cite{Shakura:1973, Cheng:2019}, and there are no mechanisms that are able to reproduce a steeper profile without invoking a strong contribution from the emission of a young ($<50$ Myr) stellar population \cite{Tacchella:2022, Cullen:2023, Topping:2023}. 

The best-fitting SED models presented in Table~\ref{tab:summary} indicate a modest but non-zero amount of reddening by dust, with an $A_V$ of 0.2--0.3 mag for both galaxies.
These results are in agreement with recent models proposed to explain the presence of luminous galaxies such as JADES-GS-z14-0 and JADES-GS-z14-1 at early times \cite{Ferrara:2023, Donnan:2023}. Such models expect that $z>10$ galaxies have lower dust content in the interstellar medium than equal-mass galaxies at lower redshifts, despite the rapid ($\sim10$~Myr) dust enrichment from supernovae \cite{Schneider:2023}. Indeed, if our massive $z\sim14$ galaxies had a stellar-to-dust ratio of about 0.002, similar to those observed in $z\sim6$  galaxies \cite{Witstok:2023}, the dust attenuation would be a factor at least 4 times higher ($A_\mathrm{V}>1$~mag), due to their compact UV size, than what is observed \cite{Ferrara:2023}.
The moderate dust attenuation in our galaxies can be explained by different scenarios: 1) a large amount of dust distributed on large scales due to galactic outflows, reducing the observed dust attenuation \cite{Ziparo:2023, Ferrara:2023};  2) a different dust composition \cite{Markov:2024} and so dust mass absorption coefficient; 3) a high destruction rate of dust grains due to shock waves generated by supernovae explosions  \cite{Schneider:2023}. Independently from the proposed scenarios, our observations indicate that the properties of galaxies appear to change rapidly in only 600~Myr (i.e. from $z=14$ to $z=6$).

In conclusion, the presented spectroscopic observations of JADES-GS-z14-0 and JADES-GS-z14-1 confirm that bright and massive galaxies existed already only 300 Myr after the Big Bang, and their number density \cite{Robertson:2023a} is more than ten times higher than extrapolations based on pre-JWST observations \cite{Bouwens:2021}. The morphology and UV slope analysis help rule out a significant AGN contribution for either galaxy. 
Other potential explanations, such as dust content \cite{Ferrara:2023}, star-formation processes \cite{Dekel:2023, Donnan:2024}, and a top-heavy initial mass function \cite{Yung:2023, Woodrum:2023}, must be investigated to explain the excess of luminous galaxies in the early Universe.

In the context of future observations, we stress that JADES-GS-z14-0 is unexpectedly and remarkably luminous. The spectroscopic confirmation of this source implies the existence of many similar galaxies -- particularly when considering the relatively small survey area of JADES. Galaxies like this are sufficiently luminous for follow-up observations with ALMA and MIRI, promising to open the view to the rest-frame optical and far-infrared at Cosmic Dawn, the period where the first galaxies were born.

\newpage

\renewcommand{\figurename}{Extended Data Fig.}
\renewcommand{\tablename}{Extended Data Table}
\setcounter{figure}{0}
\setcounter{table}{0}

\section*{Methods}\label{Method}

\subsubsection*{Cosmology model and definitions}
Throughout this study, we adopt the following cosmological parameters: $H_0$ = 70 km s$^{-1}$ Mpc$^{-1}$, $\Omega_\mathrm{m}$ = 0.3 and  $\Omega_\mathrm{\Lambda}$ = 0.7.  1 arcsec at $z = 14$ corresponds to a  physical scale of 3.268~kpc. All magnitudes are presented in the AB systems, and the term SFR$_\mathrm{10Myr}$ refers to the star formation rate averaged over the past 10 Myr. Equivalent widths of emission lines are quoted in the rest-frame. The absolute UV magnitude is estimated at the rest-frame wavelength of 1500~\AA.

\subsubsection*{NIRSpec observations and data reduction}
\label{sec:Nirspec}

\begin{table}[h!]
    \centering
    \caption{Exposure times.}
    \begin{tabular}{l c c c}
    \hline
    \hline

& JADES-GS-z14-0& JADES-GS-z14-1 & JADES-GS-53.10763-27.86014   \\
PRISM/CLEAR  &  33612s & 67225s&  67225s\\
G140M/F070LP  & 8403s & 16806s& 16806s \\
G235M/F170LP  & 8403s&16806s & 16806s\\
G395M/F290LP  & 8403s& 16806s&  16806s\\
G395H/F290LP  & 8403s & 16806s & 16806s \\

    \hline
    \end{tabular}
    \label{tab:obs}
\end{table}

The	NIRSpec data used in this work are part of the Guaranteed Time Observations program ID 1287. 
The microshutter array (MSA) mask was designed with the eMPT software\cite{Bonaventura:2023} and proceeded using the same method as described in Bunker et al.\cite{Bunker:2023} and D'Eugenio et al.\cite{DEugenio:2024}. The NIRSpec pointing was optimized for 6/7 of the highest priority targets in the catalog, which include the galaxies analyzed in this work. The eMPT software guarantees that all galaxies are located within $\sim90$ mas from the center of the shutter in the dispersion direction and within $\sim220$ mas in the spatial direction. We note that given the number of targets, not all highest priority galaxies are placed at the center on the shutter where the slit-losses effect is minimal.

The observations were carried out between January 10 and 12,	2024. Three consecutive visits were scheduled for the program, but the  visit 2 was not performed because of a loss of lock on the guide star. The two obtained visits, 1 and 3, differed in their pointing by $<1$~arcsec but they have with the same position angle. The MSA configurations were designed to place the highest priority targets in the same position within the shutter in all visits.  JADES-GS-z14-1 was observed in both visits, while JADES-GS-z14-0 was observed only in the visit 1, as the visit 3 was set up to observe the nearby low-redshift galaxy to exclude contamination.

The disperser-filter configurations employed in the program were PRISM/CLEAR, G140M/F070LP, G235M/F170LP, G395M/F290LP and G395H/F290LP. The first four spectral configurations provided spectroscopic data with spectral resolving power of $\mathrm{R=\Delta\lambda/\lambda\sim100}$ and $\mathrm{R\sim1000}$ in the wavelength	range between 0.6~$\mu$m and 5.2~$\mu$m. The G395H/F290LP disperser-filter configuration covered the wavelength range 2.87–5.27~$\mu$m  with spectral resolving power of $\mathrm{R\sim2700}$.

For the PRISM/CLEAR configuration, four sequences	of three nodded exposures were used for each pointing, while one sequence of three nodded exposures was used for the spectral configuration of the gratings. Each nodded exposure sequence consisted of six integrations of 19 groups in NRSIRS2 readout mode\citep{Rauscher:2012},  resulting in an exposure time of 8403.2 seconds.  The total exposure times for each target are reported in Extended Data Table~\ref{tab:obs}.

We made use of the NIRSpec GTO pipeline (Carniani et al. in prep.) to process the data. The pipeline was developed by the ESA NIRSpec Science Operations Team and the NIRSpec GTO Team. A general overview of the data processing is reported in Bunker et al. \cite{Bunker:2023} and D'Eugenio et al.\cite{DEugenio:2024}. To optimize the signal-to-noise ratio of the data, we used the 1D spectra extracted from an aperture of 3 pixels, corresponding to 0.3~arcsec, located at the target position in the 2D spectra.

\begin{figure}
    \centering
    \includegraphics[width=0.5\linewidth]{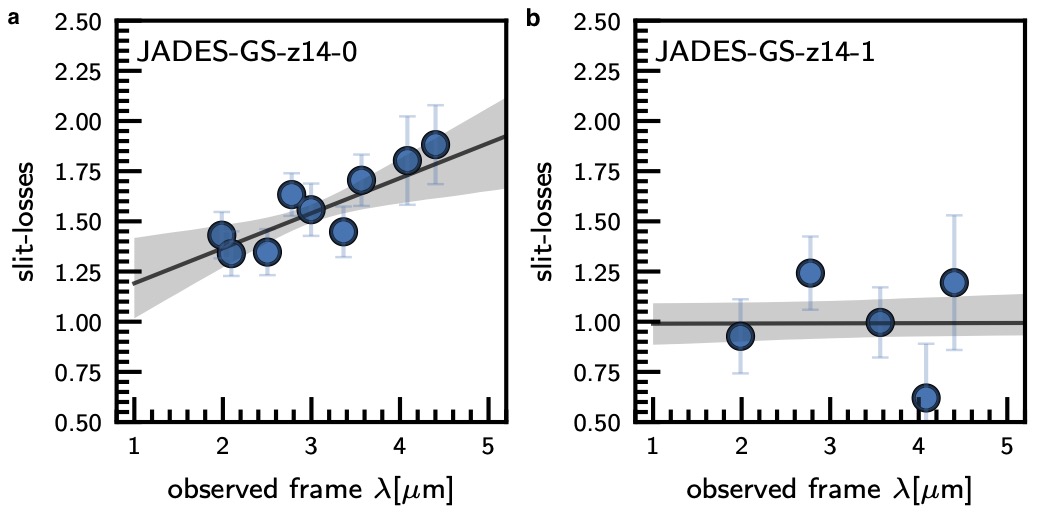}
    
    \caption{{\bf NIRSpec slit-losses. a-b,} Slit-losses as a function of wavelength for JADES-GS-z14-0 ({\bf a}) and JADES-GS-z14-1 ({\bf b}). Slit-losses are estimated by comparing the NIRSpec continuum level and NIRCam fluxes. The first-order polynomial best fits and uncertainties are reported as black lines and gray-shaded regions.}

    \label{fig:slitlosses}
\end{figure}

The pipeline applies a wavelength-dependent slit-loss correction to the measured flux, based on the position of the source inside the micro-shutter and assuming a point-source geometry.
To verify the quality of this correction, we compare the pipeline-corrected fluxes to the NIRCam photometric measurements.
For JADES-GS-z14-0, which is spatially extended, we used the NIRCam measurements derived with \texttt{ForcePho} (see Sec.~Morphological analysis. We found that $\sim30\%$ ($\sim50\%$) of the total flux is lost at 2(5)~$\mu$m in the NIRSpec data (Extended Data Fig.~\ref{fig:slitlosses}) . We employed a first-order polynomial $\alpha_1+\alpha_0\lambda$ to fit the slit losses as a function of wavelength ($\lambda $) and found $\alpha_0=0.18\pm0.11~\mathrm{\mu m^{-1}}$, $\alpha_1=1.0\pm0.2$.

For JADES-GS-z14-1, the NIRCam fluxes inferred from an aperture of 0.2 arcsec are consistent with the NIRSpec spectrum, indicating that the slit-loss correction applied by the GTO pipeline is sufficient to recover the total light of this compact source (Extended Data Fig.~\ref{fig:slitlosses}). In this case we estimated $\alpha_0=0.001\pm0.025~\mathrm{\mu m^{-1}}$, $\alpha_1=0.98\pm0.12$.

\subsubsection*{Imaging data}\label{sec:nircam}

\begin{table}
    \caption{NIRCam and MIRI fluxes of JADES-GS-z14-0 and JADES-GS-z14-1.}
    \centering
    
    \begin{tabular}{lcc}
        \hline
        \hline
         & JADES-GS-z14-0 & JADES-GS-z14-1\\
         & [nJy] & [nJy] \\
        \hline
        F090W & $-2.1\pm 0.6$ & $3.3\pm1.1$ \\
        F115W & $-0.8\pm 0.4$ & $-0.4\pm0.9$ \\
        F150W & $ 1.2\pm 0.4$ & $0.7\pm 0.9$ \\
        F162M & $-1.5\pm 0.9$ &  \\
        F182M & $13.9\pm 0.4$ & \\
        F200W & $34.8\pm 0.5$ & $7.5\pm0.7$ \\
        F210M & $46.5\pm 0.5$ & \\
        F250M & $47.2\pm 0.5$ & \\
        F277W & $55.1\pm 0.6$ & $10.1\pm0.3$ \\
        F300M & $49.8\pm 0.5$ & \\
        F335M & $43.4\pm 0.5$ & $4.5\pm0.8$\\
        F356W & $47.3\pm 0.5$ & $7.6\pm0.3$ \\
        F410M & $46.1\pm 0.8$ & $4.4\pm0.6$ \\
        F444W & $46.9\pm 0.6$ & $8.0\pm0.4$ \\
        F770W & $74.4\pm 5.6$$^\dagger$ & \\ 
        \hline

    \end{tabular}
    {\bf Note:} $^\dagger$ from Helton et al.\cite{Helton:2024}.
    \label{tab:nircam}
\end{table}

Photometry for the three candidate $z\sim14$ galaxies studied in this work was taken from JWST/NIRCam imaging catalogs of JADES \cite{Rieke:2023}, with supplemental imaging data from the First Reionization Epoch Spectroscopic COmplete Survey (FRESCO; \citep{Oesch:2023}) and JADES Origins Field (JOF; \citep{Eisenstein:2023a}) programs. These data were reduced together following the procedure outlined in Eisenstein et al.\citep{Eisenstein:2023}. The resulting mosaics include both observations taken in late 2022 as well as additional JADES observations taken in late 2023, and reach 5$\sigma$ observational depths of $2.4$ nJy in F200W using a $0.2^{\prime\prime}$ diameter aperture. We present NIRCam thumbnails centered on JADES-GS-z14-0 and JADES-GS-z14-1 in the top panels of Fig. \ref{fig:spectra}.

The sources were initially selected from the modelling of the photometry presented in Hainline et al.\cite{Hainline:2023a}. Two of the sources we explore, JADES-GS-z14-1 (JADES-GS-53.07427-27.88592 in Hainline et al.\cite{Hainline:2023a}), and the faintest galaxy described herein (JADES-GS-53.10763-27.86014) were part of the primary sample of $z > 8$ galaxies in Hainline et al.\cite{Hainline:2023a}, with photometric redshifts of $z_{\mathrm{phot}} = 14.36^{+0.82}_{-1.4}$ and $z_\mathrm{phot} = 14.44^{+0.97}_{-1.2}$, respectively. JADES-GS-z14-0 (JADES-GS-53.08294-27.85563) was presented in \cite{Hainline:2023a} (at $z_{\mathrm{phot}} = 14.51^{+0.27}_{-0.28}$) but initially rejected in that study due to the morphology, brightness, and the proximity of the source to the neighboring galaxy with photometric evidence of a Balmer break at $\sim1.7\mu$m. 

In October 2023, JADES-GS-z14-0 and JADES-GS-53.10763-27.86014  were also observed as part of the JOF program \cite{Eisenstein:2023a}, which included a NIRCam pointing of 104 hours of total exposure spread between six medium-band filters (F162M, F182M, F210M, F250M, F300M, F335M). These filters were chosen to help refine high-redshift galaxy selection in this ultra-deep region of the JADES footprint. NIRCam medium bands can be used to trace the galaxy stellar continuum and aid in rejecting sources at low redshift with strong emission lines that have similar wide-filter colors to high-redshift galaxies \cite{Naidu:2022, Zavala:2023, Arrabal-Haro:2023a}. In Robertson et al.\cite{Robertson:2023a}, the authors used JOF photometry to select a sample of nine candidate galaxies at $z = 11.5 - 15$, including JADES-GS-z14-0. The additional medium-band observations for this source had a best-fit photometric redshift of $z_{\mathrm{phot}} = 14.39^{+0.23}_{-0.09}$, and fits at $z < 7$ were effectively ruled out because of the lack of flux observed shortward of the Lyman-$\alpha$ break, the strength of the break implied by the F182M -- F210M color, and the F250M flux tracing the UV continuum. The authors also estimated a UV slope of $-2.40\pm0.12$ and a size of $260\pm6$~pc, which are consistent with those inferred in this study.
Robertson et al.\cite{Robertson:2023a} also presented the evolution of the UV luminosity function and cosmic star formation rate density at $z > 14$ inferred from observations of JADES-GS-z14-0, and we refer the reader there for more details.

For the present analysis, we fit JADES-GS-z14-0 using \texttt{ForcePho} (Johnson~B. et~al., in~prep.) in order to properly disentangle the flux of this source from the neighbor. For JADES-GS-z14-1, as this source was isolated and much more compact, we extracted fluxes using aperture photometry with an $0.2^{\prime\prime}$ aperture, and applied an aperture correction assuming a point source.  
In a companion paper \citep{Helton:2024}, our team presents JWST/MIRI photometry of JADES-GS-z14-0 from ultra-deep 43 hr F770W imaging from Program ID 1180.  This measures F770W to be $74\pm 5$ nJy, mildly above the 3--5~$\mu$m photometry of $\sim$47~nJy, likely due to the presence of strong emission lines in F770W.  Helton et al.\cite{Helton:2024} also discuss the implications of this rest-optical finding.

\subsubsection*{The third candidate JADES-GS-53.10763-27.86014}

\begin{figure}
    \centering
    \includegraphics[width=1\linewidth]{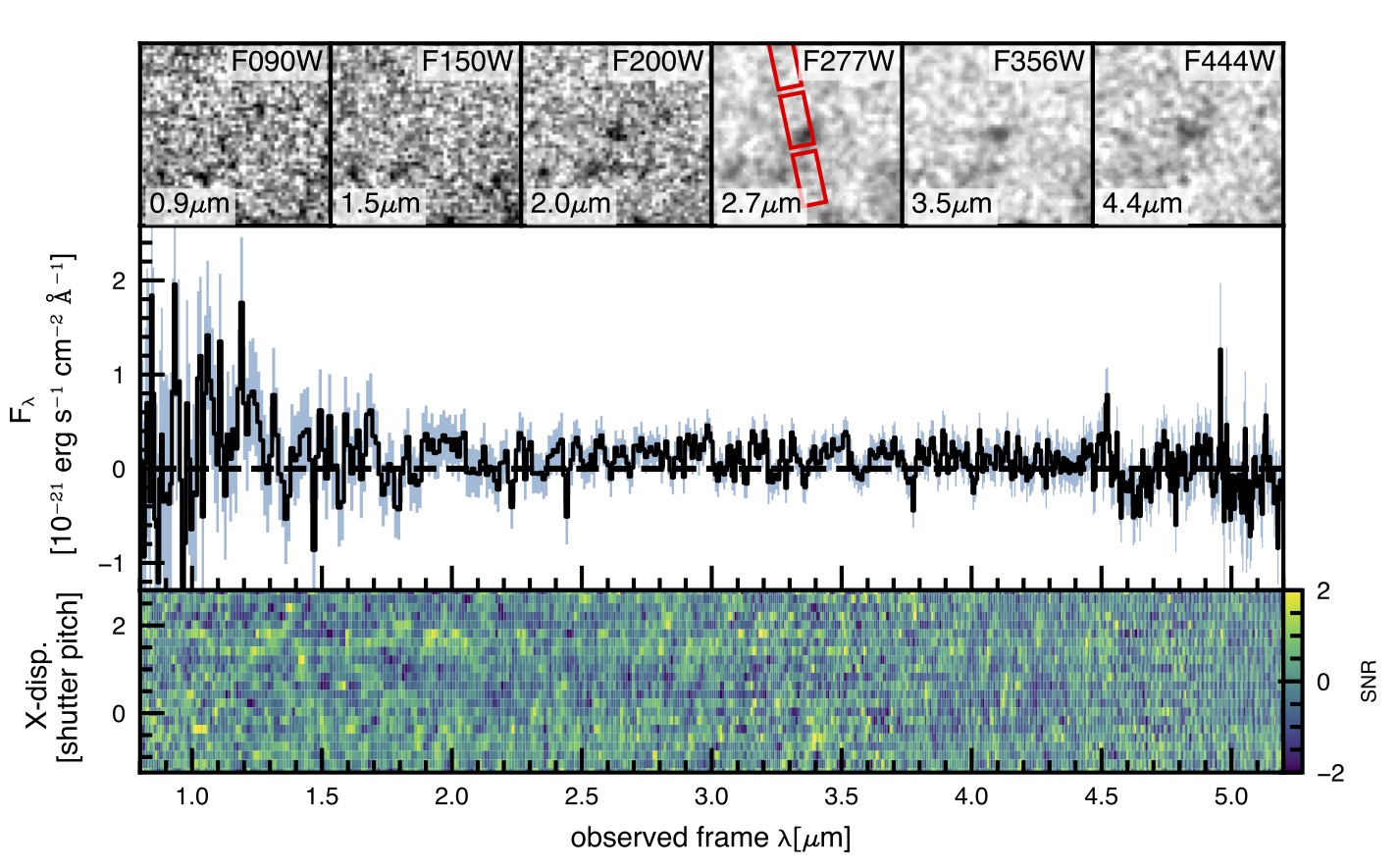}
    \caption{1D (middle) and 2D (bottom) PRISM/CLEAR spectra of JADES-GS-53.10763-27.86014, the third candidate $z\sim14$ galaxy observed with NIRSpec in the observing program 1287. Due to the low signal-to-noise ratio, no redshift can be inferred from this spectrum. The top row shows cutouts of NIRCam F090W, F150W, F200W, F277W, F356W, and F444W images. NIRSpec micro-shutter 3-slitlet array aperture is reported in red in the F277W image. }

    \label{fig:55733}
\end{figure}

The candidate $z\sim14$ galaxy JADES-GS-53.10763-27.86014 identified by Hainline et al.\cite{Hainline:2023a} and Robertson et al.\cite{Robertson:2023a}. The NIRCam images display a clear dropout in bluer filters, yielding a photometric redshift of $z_\mathrm{phot} = 14.63^{+0.06}_{-0.75}$ \cite{Robertson:2023a}. The target was observed in visits 1 and 3 of the NIRSpec program by using the same shutter position in both visits. Extended Data Fig.~\ref{fig:55733} illustrates the 1D and 2D spectra of the galaxy. Only a faint continuum emission is barely detected in the NIRSpec spectrum with a significance level less than $1\sigma$. The signal is not sufficient to confirm or rule out the photometric redshift determined by NIRCam images. We believe that the NIRSpec slit-losses contribute to the low signal-to-noise ratio of the data. The target is located at the edge of the shutter, and, despite its compact size, we expect that about 20\% of the light is lost at 2~$\mu$m and 35\% at 5~$\mu$m. The slit losses are two times higher than those of JADES-GS-z14-1, which is also 1.6 times more luminous than  JADES-GS-53.10763-27.86014. In conclusion, the low sensitivity of these observations does not allow us to confirm or rule out the photometric redshift for this target.

\subsubsection*{The low-redshift galaxy close to JADES-GS-z14-0}

\begin{figure}
    \centering
    \includegraphics[width=1\linewidth]{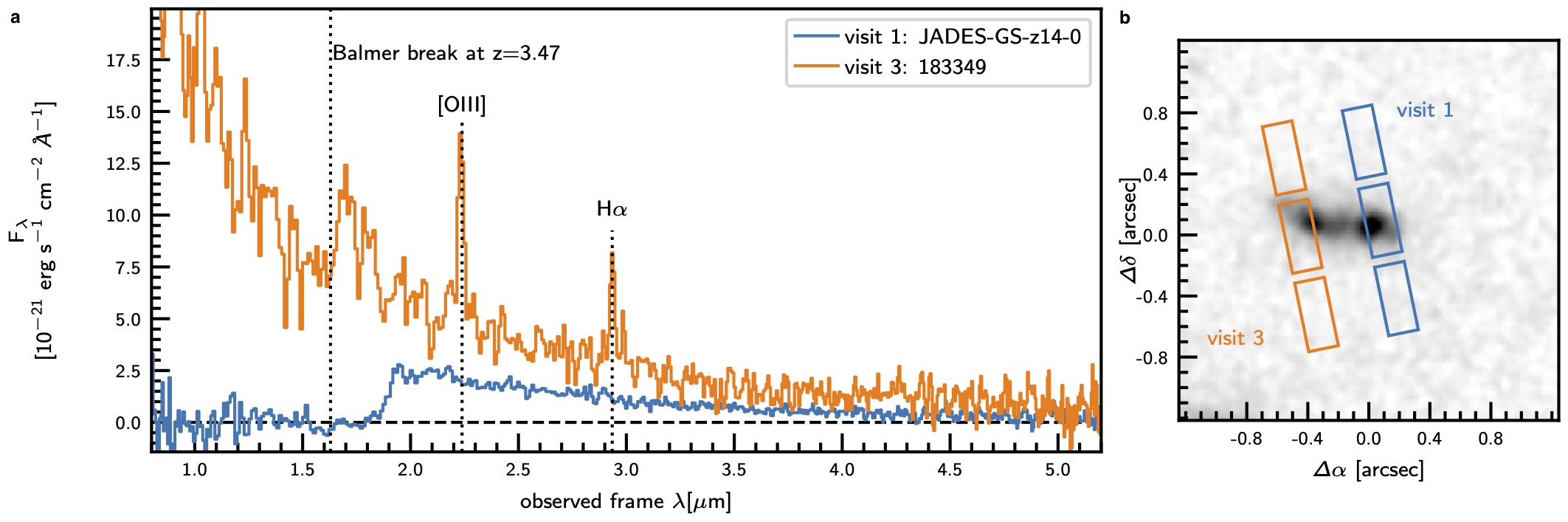} 
    \caption{{\bf Low-redshift neighbouring galaxy.} {\bf a,} Spectra of JADES-GS-z14-0 (blue) and of the target 183349 (orange). This latter is within 0.4~arcsec from JADES-GS-z14-0, as shown in the NIRCam F277W cutout in the right panel ({\bf b}). The MSA slitlets used for the two targets are overplotted in blue and orange, respectively.
}

    \label{fig:low_z_galaxy}
\end{figure}


The identification of the neighboring galaxy with NIRCam ID 183349 at 0.4 arcsec from JADES-GS-z14-0 initially raised several doubts about the photometric redshift of the high-redshift target as the potential Lyman-$\alpha$ break for this object could be a Balmer break if these two sources are associated at similar redshifts.
Therefore we dedicated the visit 3 of the NIRSpec program 1287 to observe the neighboring galaxy and assess any possible contamination and constrain the gravitational lensing effect.  Extended Data Figure~\ref{fig:low_z_galaxy} shows the spectrum of the target 183349.  The doublet [OIII]$\lambda\lambda4959,5007$ and H$\alpha$ emission lines are detected with a high level of significance in both prism and grating spectra, yielding a secure spectroscopic redshift of $z=3.475$ (in agreement with the photometric redshift $z_\mathrm{phot}=3.4\pm0.2$ from Hainline et al.\cite{Hainline:2023a}). The spectrum also reveals a clear Balmer break feature at $\sim1.6~\mu$m.  Therefore, we can rule out the drop at $\sim1.9~\mu$m observed in JADES-GS-z14-0 being due to the contamination of the neighboring galaxy. Finally, 183349 has no bright emission lines at observed wavelengths at 2.89~$\mu$m, where we detect tentative CIII] emission in JADES-GS-z14-0. We can thus rule out that the tentative CIII] is due to contamination from 183349.

\begin{figure}
    \centering
    \includegraphics[width=0.5\linewidth]{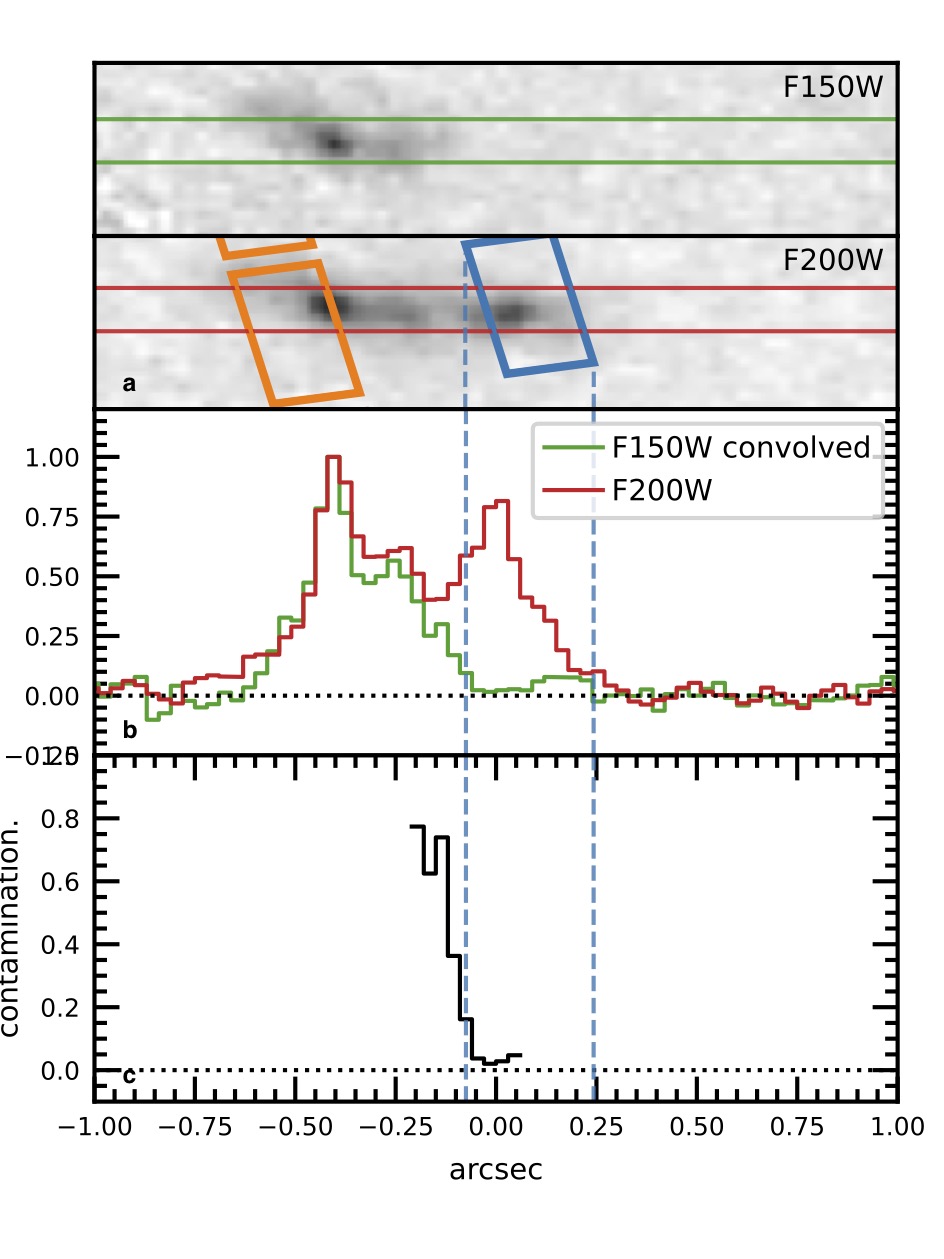}
    
    \caption{{\bf Light profile of the foreground galaxy. a--b,}. F150W ({\bf a}) and F200W  ({\bf b}) NIRCam images of JADES-GS-z14-0. The NIRSpec micro-shutter slits of visits 1 and 3 are illustrated in blue and orange, respectively. F150W image is smoothed to the same angular resolution as F200W NIRCam image. The green and red lines show the box-car extraction regions from which we obtained the light profiles reported with the same colors in the third panel. The profiles are normalized to the peak of the foreground galaxy. {\bf c,} Ratio between the F200W and F150W light profiles in the region where the two galaxies spatially overlap. This ratio quantifies the contamination of the foreground galaxy to JADES-GS-z14-0. Vertical blue dashed line shows the projection of the  NIRSpec micro-shutter slit into the radial profile.}
    \label{fig:contamination}
\end{figure}

As the foreground galaxy might contaminate the spectrum of the JADES-GS-z14-0, we analyze the surface brightness profile of the two galaxies.  Extended Data Figure~\ref{fig:contamination} shows the light profiles from the F150W and F200W NIRCam images extracted from a slit-oriented East-West and as large as 0.15~arcsec so that the slit includes both galaxies. JADES-GS-z14-0 is absent in the  F150W NIRCam image, and thus, we used the F150W profile to quantify the contamination. Before the extraction, F150W NIRCam was smoothed to the same angular resolution of F200W data. Extended Data  Figure~\ref{fig:contamination} reports the light profiles in the two filters normalized to the peak at the location of the foreground source (i.e., $\sim0.4$~arcsec from the JADES-GS-z14-0). Assuming that the surface brightness profile of the foreground galaxy in the F200W image is similar to that at F150W wavelengths, we estimated a contamination of less than 10\% at the location of JADES-GS-z14-0. The last panel of Extended Data Fig.~\ref{fig:contamination} indeed illustrates the ratio between the light profile in F200W filter and that in F150W from --0.2~arcsec (i.e., --0.6255~kpc) to 0.0~arcsec with respect to the center of JADES-GS-z14-0. This spatial range corresponds to the region in which the light of the two galaxies might overlap. The contamination is of the order of 70\% at 650~pc from the center of JADES-GS-z14-0 and drops rapidly to less than 20\% at 350 parsecs from the galaxy. As the top-left edge of the NIRSpec shutter is located at --0.08~arcsec from the center of JADES-GS-z14-0, we concluded that the contamination of the light of the foreground galaxy is negligible in the NIRSpec spectrum. Therefore, the contamination of the low-$z$ galaxy on the NIRSpec spectrum is lower than 10\%.

We have also verified that the magnification provided by the foreground galaxies to JADES-GS-z14-0 is limited ($\mu<1.2$).
We use the software \texttt{lenstool} \cite{Jullo:2007} to construct lens model of ID 183349 and another galaxy, JADES-GS-53.08324-27.85619 (ID 182698; $z_\mathrm{phot} \sim 2.04$) that is 2.2\,arcsec from JADES-GS-z14-0.
Based on HST/ACS and JWST/NIRCam SED, we infer stellar mass $\log (M_\mathrm{star}/\mathrm{M}_\odot) = 8.7 \pm 0.1$ and $9.7 \pm 0.1$ for ID 183349 and 182698, respectively. 
We then derived integrated velocity dispersions of 53 and 100 km\,s$^{-1}$ assuming the stellar-mass Tully-Fisher Relation measured at $z\sim2.3$ \cite{Uebler:2017}. 
Assuming a singular isothermal spherical distribution of matter in these two foreground potentials, we derive a modest lensing magnification factor of $\mu=1.17$ at the location of JADES-GS-z14-0. 
Such a magnification factor is corrected for when we derived the physical properties, e.g., luminosities and masses. 

\subsubsection*{Redshift determination}

\begin{figure}
    \centering
    \includegraphics[width=0.66\linewidth]{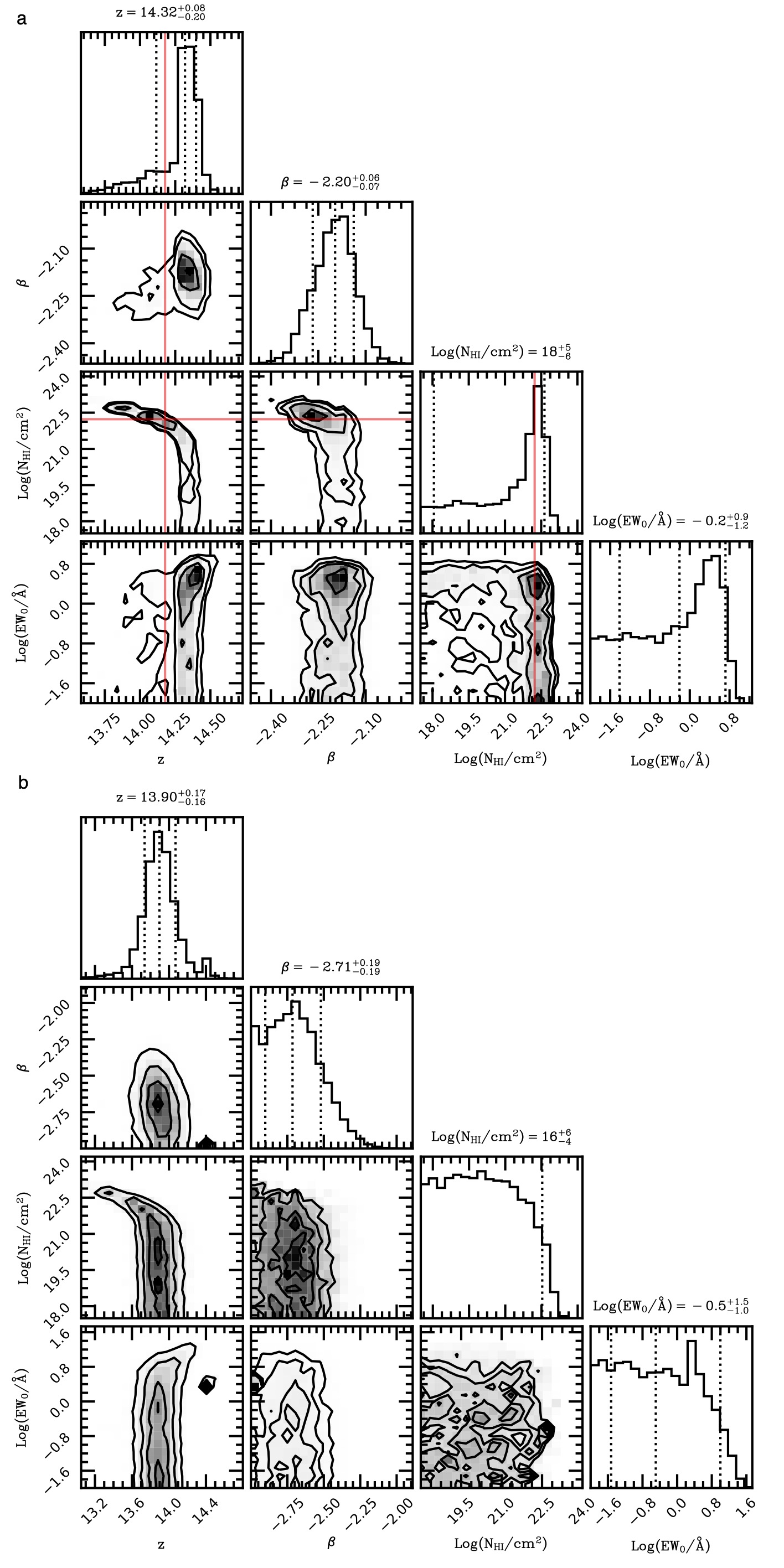}
    \caption{{\bf Redshift determination. a--b,} Results of the spectral fitting of JADES-GS-z14-0 ({\bf a}) and JADES-GS-z14-1 ({\bf b})  while	varying	the	redshift ($z$), UV slope ($f_\lambda\propto\lambda^{\beta}$), fraction of	neutral	hydrogen in	the	IGM ($X_\mathrm{HI}$), hydrogen column density of the additional damped Lyman-$\alpha$ absorption (log$(N_\mathrm{HI}/\mathrm{cm^{-2}})$), size of the ionized bubble ($R_\mathrm{ion}$),  equivalent width (log$(EW_\mathrm{0}/\text{\AA})$) and velocity shift ($\Delta v$) of Lyman-$\alpha$ line. The corner plots show the posterior distribution functions and the covariance between the parameters. The posterior distribution of $X_\mathrm{HI}$, $R_\mathrm{ion}$, and $\Delta v$ is not illustrated because they are flat. The dotted lines indicate 16th, 50th, and 84th percentile of the marginalized posterior distributions.
    The red lines in the corner plot of JADES-GS-z14-0 represent the estimates of redshift and log$(N_\mathrm{HI}/\mathrm{cm^{-2}})$ based on the potential CIII]$\lambda1909$ emission line.}
    \label{fig:redshift_long}
\end{figure}

The photometric redshifts of JADES-GS-z14-0 and JADES-GS-z14-1 are $z_a=14.39^{+0.23}_{-0.09}$\cite{Robertson:2023a} and $z_{\mathrm{phot}} = 14.36^{+0.82}_{-1.4}$ \citep{Hainline:2023a}, respectively . The strong Lyman breaks (flux ratio
120 between 1.90 - 2.1 $\mu$m and 1.5 - 1.8 $\mu$m higher than 9) observed by NIRSpec in both JADES-GS-z14-0 and JADES-GS-z14-1 confirm both galaxies to be at a redshift of about $\sim14$.

Recent studies have shown that the profile of the Ly$\alpha$ spectral break does not depend only on intergalactic medium absorption but can also be modulated by: a) neutral gas in the galaxy or in the surrounding medium \cite{Heintz:2023, DEugenio:2023, Heintz:2024}; b) the presence of a local ionized bubble \cite{Saxena:2023, Witstok:2024}; c) Ly$\alpha$  line emission which would enhance the flux of spectral channels containing the line in the low-resolution data \cite{Keating:2023}. Therefore, to determine the spectroscopic redshift, we model the continuum emission with a power-law function ($f_\lambda\propto\lambda^{\beta}$), which well reproduces the rest-frame UV continuum emission in galaxies with young stellar populations \cite{Schaerer:2005,Tacchella:2022, Cullen:2023, Topping:2023}, and absorption of neutral hydrogen following the prescriptions discussed in Witstok et al.\cite{Witstok:2024} and Hainline et al.\cite{Hainline:2024}.  The IGM transmission is modeled following Mason \& Gronke \cite{Mason:2020} and depends on two free parameters: the global neutral gas fraction ($x_\mathrm{HI}$) and the ionized bubble size ($R_\mathrm{ion}$). We assumed a flat prior for the neutral gas fraction over the range  $x_\mathrm{HI}\in[0.95, 1]$ and a flat prior distribution for the ionized bubble size over the range $R_\mathrm{ion}\in[0.1, 1]$ proper Mpc. These are the expected values for a typical galaxy with a $M_\mathrm{UV}\sim-20$ at $z=14$ \cite{Neyer:2023}.
As the Lyman-$\alpha$ drop profile can also be caused by dense neutral gas in the circumgalactic medium and located along the line of sight (i.e., damped Lyman-$\alpha$ absorption) following D'Eugenio et al.\cite{DEugenio:2023} and  \cite{Hainline:2024}, we parametrized this additional absorption by the column density of neutral hydrogen, $\log(N_\mathrm{HI}/\mathrm{cm^{-2}})$, and assumed a flat prior  $\log(N_\mathrm{HI}/\mathrm{cm^{-2}})\in[10,28]$.  
Finally, recent studies \cite{Keating:2023, Jones:2024} have also shown that the Ly$\alpha$ emission line can modify the prism spectra and so alter the redshift measurement. 
Therefore we added a mock spectroscopically unresolved emission line in our model to represent the  Ly$\alpha$ emission. The line model was parametrized by the rest-frame equivalent width ($\log(EW_0/ \text{\AA})\in[-2,2]$) and velocity shift with respect to systemic  ($\Delta v \in[0,3000]~\mathrm{km~s^{-1}}$). This latter mimics the effects of outflows and resonant scattering on Ly$\alpha$ line emission \cite{Orsi:2012}. 

Extended Data Figure~\ref{fig:redshift_long} shows the posterior distributions of the free parameters used to fit the data of JADES-GS-z14-0 and JADES-GS-z14-1. The posteriors of the parameters $x_\mathrm{HI}$, $R_\mathrm{ion}$, and $\Delta v$ are flat for both galaxies and are not reported in the corner plot.
The best-fit redshifts are $14.32^{+0.08}_{-0.20}$ and $13.90\pm0.17$, respectively for the two targets. The profile of the posterior distributions of $\log(N_\mathrm{HI}/\mathrm{cm^{-2}})$ exclude the presence of dense DLA with $\log(N_\mathrm{HI}/\mathrm{cm^{-2}})>23.64$, but does not preclude less dense absorbing systems along the line of sight. The results also indicate that the rest-frame equivalent width of the Lyman-$\alpha$ line is lower than 10~\AA.

\subsubsection*{Emission lines}

\begin{figure}
    \centering
    \includegraphics[width=1\linewidth]{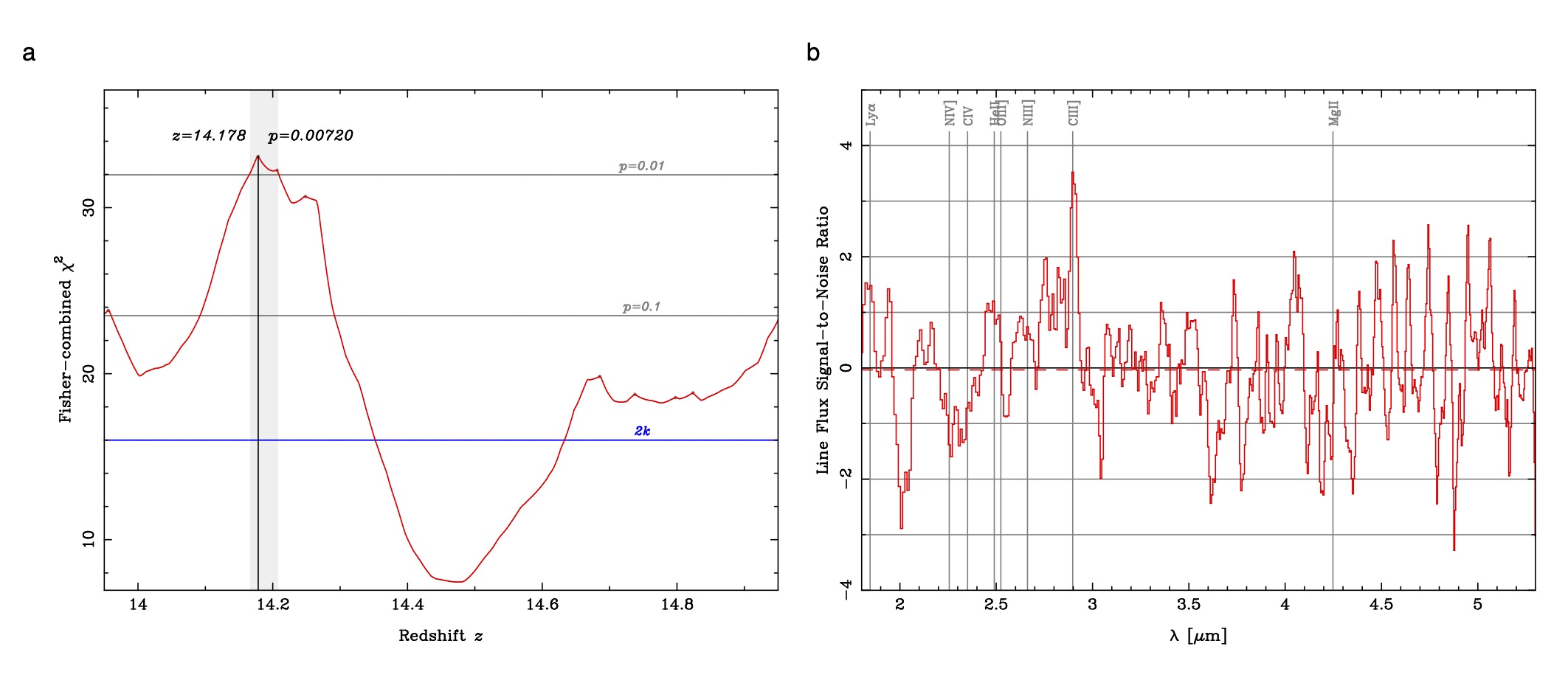}  
    \caption{{\bf Emission line identification in the prism spectrum of JADES-GS-z14-0}. {\bf a,}  Fisher-combined $\chi^2$ value as a function of redshift. The gray horizontal lines show Fisher-combined $\chi^2$ values at the combined p-values of 0.01 and 0.1. For reference, the blue line marked $2k$, where $k$ is the number of emission lines falling within the search window, reports the mean value of the $\chi^2$ statistic under the null hypothesis of no emission lines. There is a peak is at $z = 14.178$ with a  $p$-value of 0.00720. {\bf b,} Potential line flux signal-to-noise ratios and the overlay shows the locations of the emission lines searched for at the peak of the combined $p$-value. The potential CIII]$\lambda1909$ emission line has a level of significance of 3.6$\sigma$.}
    \label{fig:gsz14_0_redshift_sweep}
\end{figure}

We inspected the prism and grating 1D and 2D spectra to identify any rest-frame ultraviolet emission lines above the level of the noise in both targets.
We estimated emission line fluxes and equivalent widths from the continuum-subtracted spectra over five spectral channels (Extended Data Table~\ref{tab:emission_line}). The uncertainties on line fluxes and equivalent width were determined by repeating the measurements on a sample of 2000 spectra obtained by combining the spectra of the individual integrations with a bootstrap resampling technique.

\begin{figure}
    \centering
    \includegraphics[width=1\linewidth]{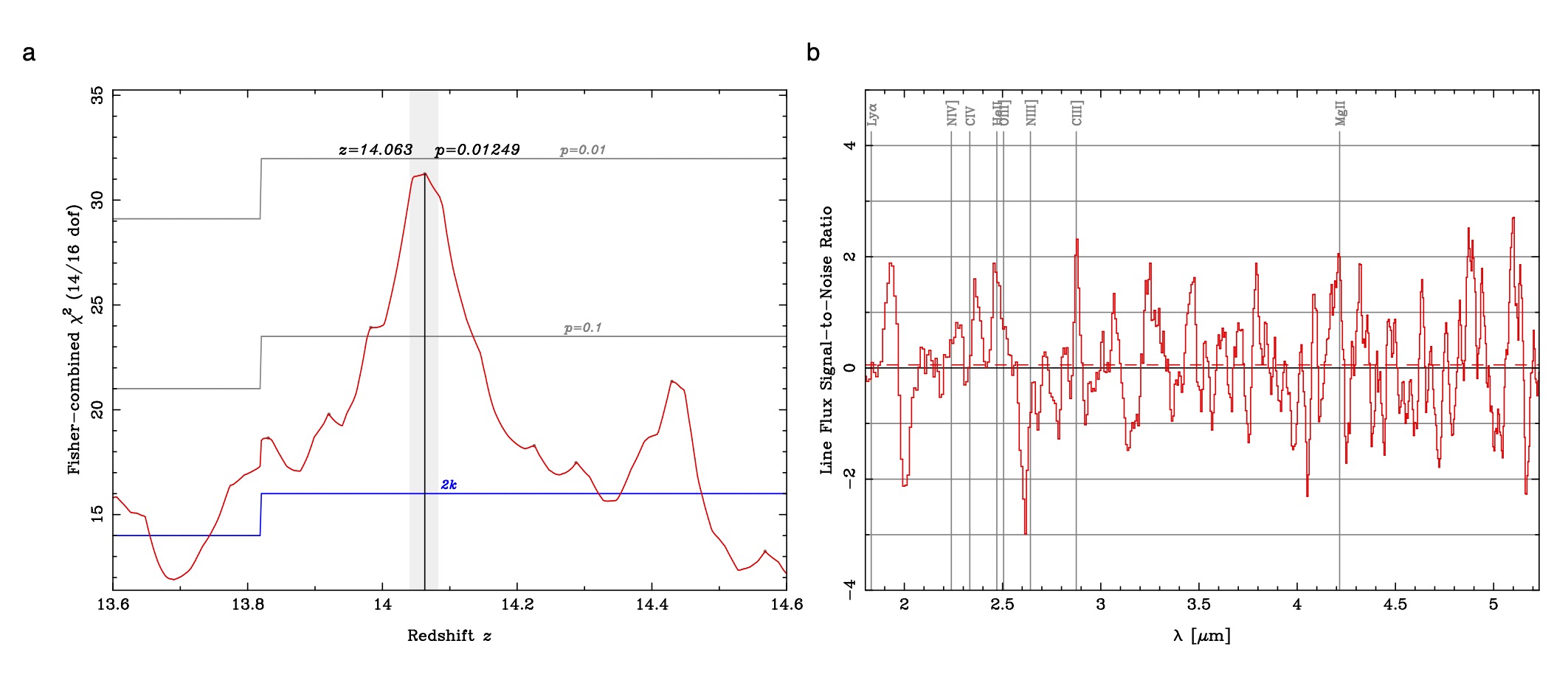}  
    \caption{{\bf Emission line identification in the prism spectrum of JADES-GS-z14-1}. {\bf a,}  Fisher-combined $\chi^2$ value as a function of redshift. The gray horizontal lines show Fisher-combined $\chi^2$ values at the combined p-values of 0.01 and 0.1. For reference, the blue line marked $2k$, where $k$ is the number of emission lines falling within the search window, reports the mean value of the $\chi^2$ statistic under the null hypothesis of no emission lines. There is a peak at $z = 14.063$ with a  $p$-value of 0.01249. {\bf b,} Potential line flux signal-to-noise ratios and the overlay shows the locations of the emission lines searched for at the peak of the combined $p$-value. There are no emission lines with a level of significance higher than 3$\sigma$.}
    \label{fig:gsz14_1_redshift_sweep}
\end{figure}

Given the uncertainties on the redshift based only on the Lyman-break, we estimated the statistical significance of a set of emission lines (see Extended Data Table~\ref{tab:emission_line}) at different redshifts (Extended Data Fig.~\ref{fig:gsz14_0_redshift_sweep} and ~\ref{fig:gsz14_1_redshift_sweep}). In particular, we inferred the one-sided $p$-value for each line at different redshifts. We then determined the combined $p$-value of the set of lines by using Fisher's method and used it to quantify the statistical significance of the spectroscopic redshifts (see details in Hainline et al.\cite{Hainline:2024}).

In the prism spectrum for JADES-GS-z14-0, we identified only a potential CIII]($\lambda\lambda1907,1909$) emission line at $z=14.178$ with a level of significance of 3.6$\sigma$ and the combined $p$-value for the inferred redshift is 0.00363. The redshift is consistent within the error with that determined from the fitting of the Lyman-break profile, but follow-up observations will be required to confirm the emission line. We note that the presence of only carbon line in the rest-frame UV spectrum is consistent with other low redshift studies concluding that CIII] might be the strongest rest-UV line after the Ly$\alpha$ line \cite{Shapley:2003, Stark:2015, Feltre:2020, Topping:2021}.

Both the upper limits on the emission lines and the tentative detection of CIII] in JADES-GS-z14-0, with a rest-frame equivalent width $EW_0=8.0\pm2.3$~\AA (Table~\ref{tab:emission_line}), are consistent with those observed in lower-luminosity galaxies at $z>10$ \cite{Curtis-Lake:2023, Hainline:2024}. 
On the other hand, if we compare JADES-GS-z14-0 with the most luminous galaxies at $z>10$, such as GN-z11 \cite{Bunker:2023a} and GHZ2 \cite{Castellano:2024}, we would expect to detect both CIII] and CIV$\lambda\lambda1548,1551$ in the prism spectra of our galaxies.
This spectral difference may be due to an extremely low metallicity ($Z<0.05~\mathrm{Z_{\odot}}$) or a large escape fraction of ionizing photons which reduces the emission by the gas in the interstellar medium \cite{Curtis-Lake:2023, Hainline:2024} or a different nature of the dominant ionizing flux \cite{Maiolino:2023}.

In the grating, we did not find any lines, and the $3\sigma$ upper limits, which were estimated by using the bootstrap resampling technique, are reported in Extended Data Table~\ref{tab:emission_line}.

\begin{table}
    \caption{Spectral measurements from prism and gratings spectra.}
    \centering
    \begin{tabular}{c|c|c||c|c||}
    \hline
         & \multicolumn{2}{c||}{JADES-GS-z14-0} &  \multicolumn{2}{c||}{JADES-GS-z14-1}\\ 
        \hline
        Emission line & Flux & $EW_{0}$&  Flux &  $EW_{0}$\\
        &  [$10^{-19}~\mathrm{ erg~s^{-1}~cm^{-2} }$] & [\AA] &  [$10^{-19}~\mathrm{ erg~s^{-1}~cm^{-2} }$] &[\AA] \\
        \hline

         & \multicolumn{2}{c||}{Prism} & \multicolumn{2}{c||}{Prism}\\ 
        \hline
        $\mathrm{ Ly}\alpha$           & $<3$   &  & $<2$ &   \\
        $\mathrm{ N IV]}\lambda$1486   & $<2$   & $<6$   & $<1.2$ & $<14$\\
        $\mathrm{ CIV}\lambda$1548     & $<2$   & $<7 $  & $<1.1$ & $<14$\\
        $\mathrm{ HeII}\lambda$1640    & $<1.9$ & $<7 $  & $<1.0$ & $<16$\\
        $\mathrm{ OIII]}\lambda$1660   & $<2$   & $<9 $  & $<1.0$ & $<15$\\
        $\mathrm{ N III]}\lambda$1750  & $<1.4$ & $<6 $ & $<0.8$ & $<14$\\
        $\mathrm{ C III]}\lambda$1908  & ($1.4\pm0.4$) & ($8.0\pm2.3$) & $<0.7$ & $<16$\\
        $\mathrm{ Mg II}\lambda$2795   & $<0.9$ & $<13$ & $<0.6$ & $<35$  \\
        \hline
        & \multicolumn{2}{c||}{Gratings} & \multicolumn{2}{c||}{Gratings}\\ 
        \hline
        $\mathrm{ Ly}\alpha$           & $<5$ & $<19$  & $<3$ & $<32$  \\
        $\mathrm{ N V }\lambda$1240    & $<4$ & $<9 $  & $<3$ & $<28$  \\
        $\mathrm{ N IV]}\lambda$1486   & $<5$ & $<8$   & $<1.6$ & $<18$\\
        $\mathrm{ CIV}\lambda$1548     & $<3$ & $<8 $  & $<1.4$ & $<16$\\
        $\mathrm{ HeII}\lambda$1640    & $<2$ & $<8 $  & $<1.3$ & $<18$\\
        $\mathrm{ OIII]}\lambda$1660   & $<2$ & $<8 $  & $<1.3$ & $<25$\\
        $\mathrm{ N III]}\lambda$1750  & $<2$ & $<10 $ & $<1.2$ & $<32$\\
        $\mathrm{ C III]}\lambda$1908  & $<3 $ & $<14$ & $<1.5$ & $<37$\\
        $\mathrm{ Mg II}\lambda$2795   & $<2$ & $<33$ & $<1.2$ & $<70$  \\

        \hline
    \end{tabular}
     {\bf Note:} Fluxes are in units of $10^{-19}~\mathrm{ erg~s^{-1}~cm^{-2} }$, rest-frame equivalent widths ($EW_0$) are in (rest-frame) \AA. We adopted 3$\sigma$ upper limits and emission line fluxes have been estimated over a resolution element of 5 spectral channels of R1000 data. 
    \label{tab:emission_line}
\end{table}

In both the prism and grating of JADES-GS-z14-1  we did not find any clear emission line feature with a level of significance higher than $3\sigma$. We identified only the potential lines of CIII]$\lambda1909$ and MgII$\lambda2795$ at $z=14.063$ with a signal-to-noise ratio of about 2. The combined $p$-value for this redshift is 0.01249, suggesting that this solution is not statistically significant.
We thus derived the upper limits on the emission lines and equivalent widths from both prism and grating spectra (Extended Data Table~\ref{tab:emission_line}). The prism spectra also reveal an absorption feature at 2$\mu$m but its significance is only 2$\sigma$. If it will be confirmed in future observations, it corresponds to CII $\lambda1335$ doublet absorption line.

\subsubsection*{Possible large-scale structure association}

\begin{figure}
    \centering
    \includegraphics[angle=-90,width=1\linewidth]{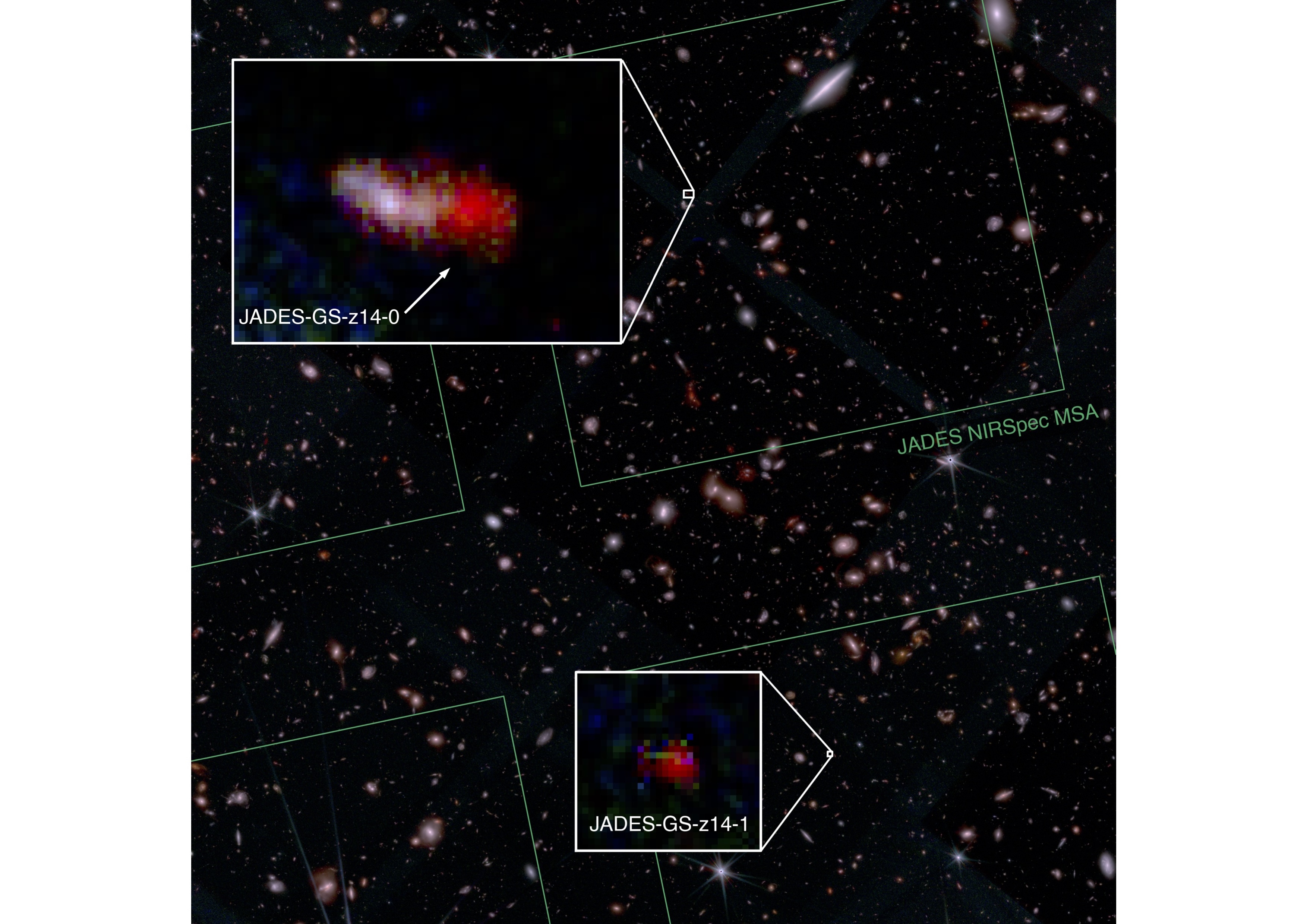}
    \caption{{\bf The sky around JADES-GS-z14-0 and JADES-z14-1.} F444W/F200W/F090W false color red/green/blue image of the JADES imaging field \cite{Robertson:2023a} (background) and  F277W/F150W/F115W false color red/green/blue thumbnail images of the two $z\sim14$ galaxies. The green outline illustrates the four quadrants of the NIRSpec micro-shutter array of the Cycle 1 NIRSpc 1287 program.} 
\label{fig:lss}
\end{figure}

JADES-GS-z14-0 and JADES-GS-z14-1 are 1.9$'$ apart on the sky (Extended Data Fig.~\ref{fig:lss}), which is 6.2 comoving Mpc at this redshift.  The third candidate completes a roughly equilateral triangle, 1.7$'$ and 2.7$'$ away from the first two, respectively.  These three galaxies form a mild angular over-abundance of the candidates from Hainline et al.\citep{Hainline:2023a}, a fact that influenced the selection of this location for the program 1287 deep NIRSpec pointing.  
The separation along the line of sight is imprecisely known, given the redshift uncertainties.  If JADES-GS-z14-0 and JADES-GS-z14-1 are separated by 0.42 in redshift, then this would be about 60 comoving Mpc along the line of sight.  However, the galaxies could be substantially closer, even potentially at the same redshift, if more unusual combinations of neutral hydrogen absorption and Ly$\alpha$ emission were present.  Narrow-line redshifts will be needed to measure this.  
As galaxies at high redshift are expected to exhibit a high clustering bias \cite{Zhang:2019,Endsley:2020}, supported by numerous findings of inhomogeneity at $z\approx 7$ \cite{Higuchi:2019,Helton:2023}, it seems likely that these galaxies are at least mildly associated in an extended large-scale structure.

\subsubsection*{SED fitting}

\begin{figure}
    \centering
    \includegraphics[width=1\linewidth]{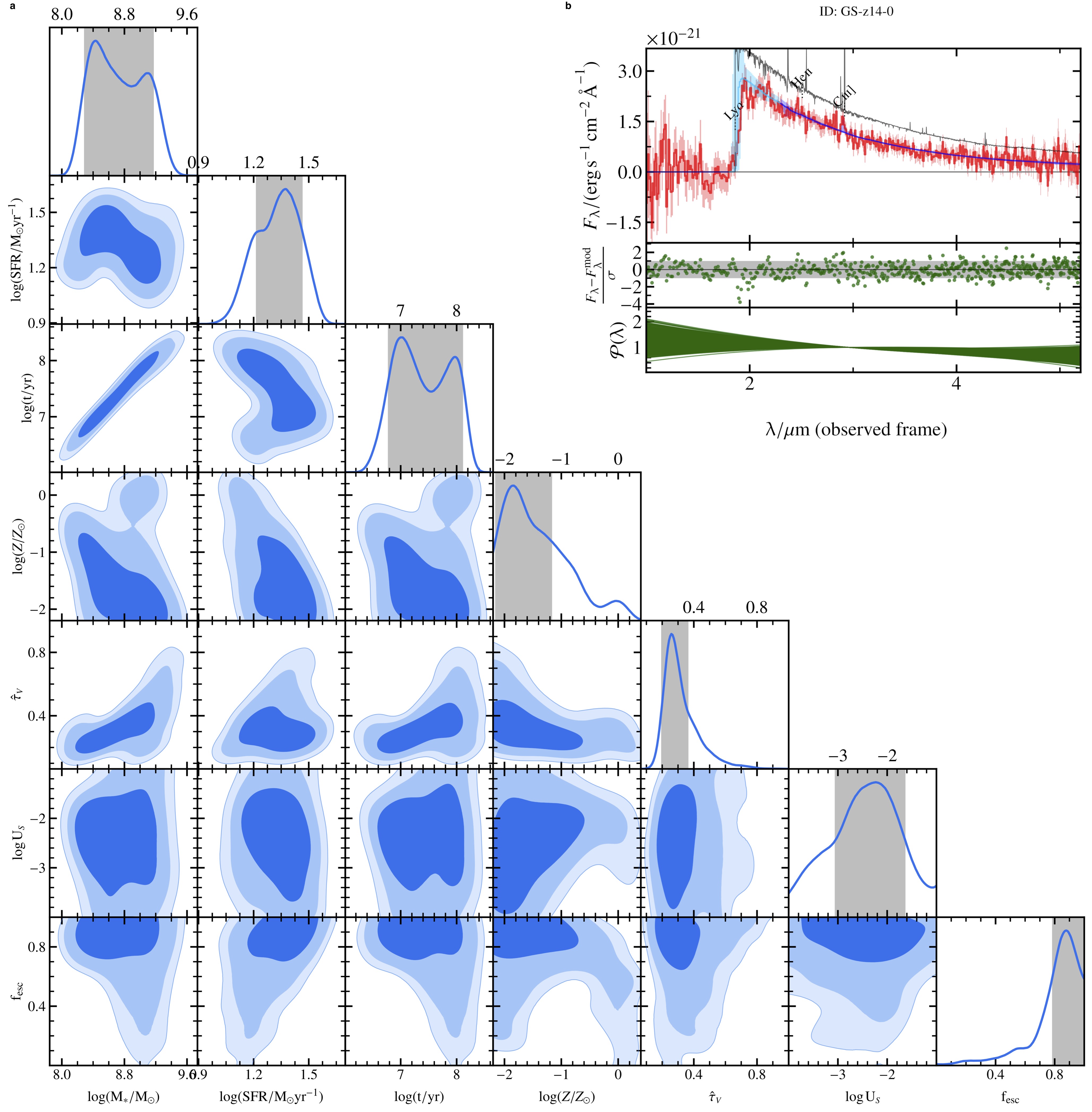}
    \caption{{\bf Beagle-based SED modeling of JADES-GS-z14-0. a,} Posterior probability distributions obtained with Beagle for our fiducial fits, along with the observed spectrum and model prediction, for JADES-GS-z14-0. From left to right, the columns show the stellar mass \Mstar, star formation rate, age of the oldest stars \age, (stellar and interstellar) metallicity \Zmetal, V-band attenuation optical depth \tauV, ionization parameter \logUs, and escape fraction of ionizing photons \fesc. The 1D (marginal) posterior distribution of each parameter is plotted along the diagonal, where the shaded gray regions represent the 1$\sigma$ credible interval. The off-diagonal panels show the 2D (joint) posterior distributions, with the shaded blue regions representing the 1, 2, and 3$\sigma$ credible intervals. {\bf b,} The observed spectrum (red line), model predictions (dark blue line), and model predictions before applying instrumental effects (i.e. line-spread function and calibration polynomial, grey line). The model predictions at $\lambda < 1450$ \AA\ are shown with a cyan line, to indicate that this region was masked during the fitting. In the central panel of the inset, we show the residuals in units of observed errors and the $\pm 1 \sigma$ region in grey, while the bottom panel indicates the calibration polynomial.} 
\label{fig:beagle_gsz14_0}
\end{figure}

We fit JADES-GS-z14-0 and JADES-GS-z14-1 following a similar approach to that of Hainline et al.\cite{Hainline:2024} and Curtis-Lake et al.\cite{Curtis-Lake:2023}. In summary, we use the \beagle\ tool \citep[BayEsian Analysis of GaLaxy sEds][]{Chevallard:2016} to fit the combined R100 NIRSpec spectra and NIRCam + MIRI photometry. We fit the entire spectrum uncorrected for slit losses and mask the rest-frame region 1150 -- 1450 \AA\ to avoid biases arising from the \Lya\ damping wing, which we do not model in \beagle. We fit the NIRCam wide-bands F090W, F115W, F150W, F200W, F277W, F356W, F444W, and the MIRI F770W band, using the values and errors reported in Extended Data Table~\ref{tab:nircam}. We adopt Gaussian priors for the redshift of the sources, centered on $z=14.32$ and with $\sigma_z=0.2$ for JADES-GS-z14-0, and centered on $z=13.9$ and with $\sigma_z=0.17$ for JADES-GS-z14-1. To account for the wavelength-dependent slit-losses of NIRSpec, which are especially important for the extended object JADES-GS-z14-0, we include in the fitting a second-order polynomial, which is applied to the model spectrum before comparing it with the observed one. This enables the model to reproduce consistently the NIRCam+MIRI photometry and uncorrected NIRSpec spectrum.
The inset in Extended Data Figure~\ref{fig:beagle_gsz14_0}, and in particular the difference between the grey (model before the polynomial correction) and blue lines (model with the applied correction), shows the importance of this correction for JADES-GS-z14-0.  We do not use the first-order polynomial correction estimated in Sec.~NIRSpec observations because we want to take into account the uncertainties associated with slit-losses correction directly in the Bayesian SED fitting.

\begin{figure}
    \centering
    \includegraphics[width=1\linewidth]{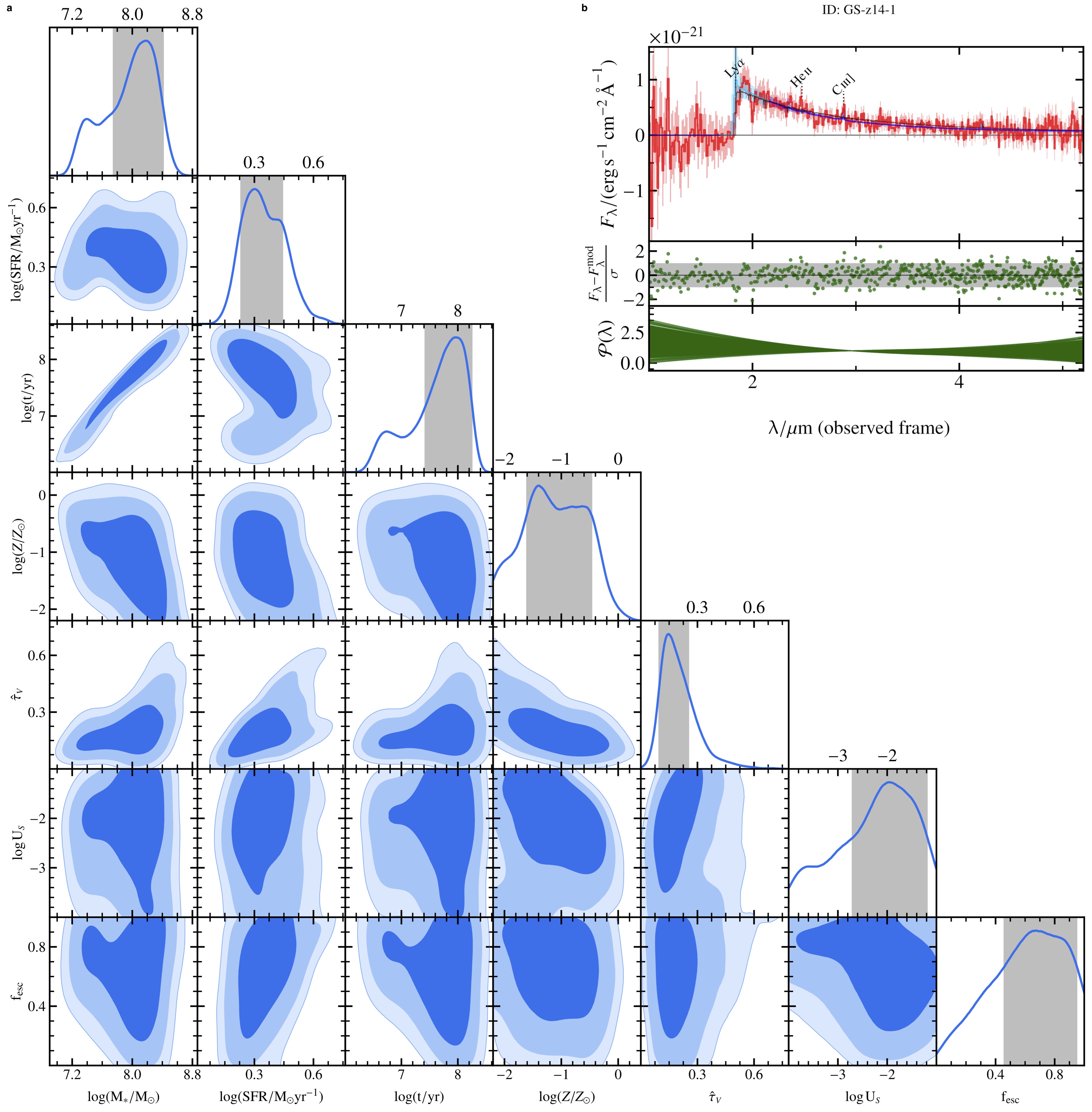}
    \caption{{\bf Beagle-based SED modeling of JADES-GS-z14-1. a,} Posterior probability distributions obtained with Beagle for our fiducial fits, along with the observed spectrum and model prediction, for JADES-GS-z14-1. From left to right, the columns show the stellar mass \Mstar, star formation rate, age of the oldest stars \age, (stellar and interstellar) metallicity \Zmetal, V-band attenuation optical depth \tauV, ionization parameter \logUs, and escape fraction of ionizing photons \fesc. The 1D (marginal) posterior distribution of each parameter is plotted along the diagonal, where the shaded gray regions represent the 1$\sigma$ credible interval. The off-diagonal panels show the 2D (joint) posterior distributions, with the shaded blue regions representing the 1, 2, and 3$\sigma$ credible intervals. {\bf b,} The observed spectrum (red line), model predictions (dark blue line), and model predictions before applying instrumental effects (i.e. line-spread function and calibration polynomial, grey line). The model predictions at $\lambda < 1450$ \AA\ are shown with a cyan line, to indicate that this region was masked during the fitting. In the central panel of the inset, we show the residuals in units of observed errors and the $\pm 1 \sigma$ region in grey, while the bottom panel indicates the calibration polynomial.}
\label{fig:beagle_gsz14_1}
\end{figure}

As in Hainline et al.\cite{Hainline:2024}, we use an updated version of the Bruzual et al. \cite{Bruzual:2003} stellar population synthesis models (see Vidal et al.\cite{Vidal:2017} for details), along with the photoionization models of Gutkin et al.\cite{Gutkin:2016}. We assume a Chabrier\cite{Chabrier:2003} initial mass function (IMF) with lower and upper mass limits of 0.1 and $300 \; M_{\odot}$, respectively. We fit three different models, in which we vary assumptions on the star formation history and escape fraction of ionizing photons. Similarly to Hainline et al.\cite{Hainline:2024}, we find that the blue UV slopes of our objects and the absence of (securely) detected emission lines, can be explained by (i) a star formation rate which suddenly drops to very low values during the last 10 Myr of star formation, but with most stars being a few tens of Myr old, as they can produce blue UV slopes. This scenario implies an unlikely fine-tuning of the star formation history of these galaxies. (ii) A metal-enriched gas (near-to-Solar metallicity), as such a large metallicity would also suppress the high-ionization UV lines (but also produce redder UV slopes). (iii) A large escape fraction of ionizing photons. This latter model is our fiducial one, since, as we extensively discussed in Hainline et al.\cite{Hainline:2024}, is the only model that can simultaneously reproduce the observed blue UV slopes and the absence of UV emission lines.

Our fiducial model is thus described by seven adjustable parameters: the total stellar mass formed \Mtot, age of the oldest stars \age, stellar metallicity \Zstar, gas ionization parameter \logUs, V-band dust attenuation optical depth \tauV (i.e., $A_{V}$/1.086), escape fraction of ionizing photons \fesc\, and redshift $z$. We assume the Charlot et al.\cite{Charlot:2000} two-component dust attenuation model, fixing the fraction of attenuation arising from stellar birth clouds to $\mu = 0.4$. We further fix the metal's depletion factor to $\xi = 0.1$. Note that in our model the stellar metallicity \Zstar\ and interstellar metallicity \Zism\ are equal, while the gas abundance of a metal further depends on the metal depletion. 

From Beagle-based SED modeling we find that the JADES-GS-z14-0, once corrected for the gravitational lensing amplification, has a stellar mass of  $\log_{10}(M_\mathrm{ star}/\mathrm{ M_\odot}) = 8.6^{+0.7}_{-0.2}$ and a star-formation rate, averaged over the last 10~Myr, of $\text{SFR}_{10} = 19\pm6~\mathrm{ M_\odot~yr^{-1}}$, resulting in a specific SFR of $\text{sSFR}=\text{SFR}/M_\mathrm{ star}=45~\mathrm{ Gyr^{-1}}$. We estimate a gas-phase metallicity of $\log{(Z/Z_\odot)} = -1.5^{+0.7}_{-0.4}$ and a dust attenuation of $A_\mathrm{ V}\sim0.3~\mathrm{ mag}$ assuming a standard nebular continuum powered by OB stars. JADES-GS-z14-1 is less massive, with $\log_{10}(M_\mathrm{ star}/\mathrm{ M_\odot}) = 8.0^{+0.4}_{-0.3}$ and  a star-formation rate of $\text{SFR}_{10} = 2.0^{+0.7}_{-0.4}~\mathrm{ M_\odot~yr^{-1}}$. We infer a $\text{sSFR}$ of about 18~$\mathrm{ Gyr^{-1}}$. For this galaxy, we estimate a metallicity of $\log({Z/Z_\odot}) = -1.1^{+0.6}_{-0.6}$ and a dust attenuation of $A_\mathrm{ V}\sim0.2~\mathrm{ mag}$.
We remind that the errors quoted here refer only to the internal statistical errors of our model with the above assumptions. Notably, the inference of stellar mass is known to be sensitive to assumptions, with variations of about 0.2 dex depending on the SED fitting code and star-formation histories allowed in the model \cite{Helton:2024}. More exotic deviations in astrophysics, such as variations in the stellar IMF \cite{Wang:2024}, could create further differences.

\subsubsection*{Morphological analysis}

To determine the extension of JADES-GS-z14-0 and JADES-GS-z14-1, we initially extracted the radial surface brightness profile from the NIRCam F277W images, which are a compromise between S/N and angular resolution for these galaxies.

We have determined the average surface brightness from concentric radial annuli centered at the position of the target and radius width of $0.03^{\prime\prime}$. For JADES-GS-z14-0 we have masked the image at Right Ascension coordinate $>53.0829890$~deg (i.e, $0.1^{\prime\prime}$ from the galaxy center) to remove the contamination from the foreground source.  Figure~\ref{fig:spatial_extension} shows the normalized surface brightness profile for both targets.
The emission of JADES-GS-z14-0 is more extended than the light profile of the star (i.e., point-like source) observed in the NIRCam field.
JADES-GS-z14-0 is also more extended than the surface brightness profile of GN-z11 \cite{Tacchella:2023} and GHZ2 \cite{Castellano:2024}. On the other hand, JADES-GS-z14-1 appears more compact and consistent with a point-like source. 

We have also modeled the morphology of the galaxies with {\tt ForcePho}, which enables us to fit the light distribution of individual exposures across all filters simultaneously while taking into account the substantial change in the NIRCam PSF with wavelength. In this case we do not need to apply any mask in the NIRCam image as {\tt ForcePho} models simultaneously the target and the neighboring galaxies returning the morphological and photometric parameters for all sources in the field.

In the F162M image, JADES-GS-z14-0 is near to the edge of or entirely missed by a number of the individual exposures from the 3215 program \cite{Eisenstein:2023a}. By modeling the individual exposures simultaneously {\tt ForcePho} avoids the correlated noise caused by mosaicing, which can be particularly difficult to quantify when the source is near a large gradient in exposure time such as JADES-GS-z14-0. We have adopted a single S\'{e}rsic profile for the $z\sim14$ galaxies and for the foreground sources within 5~arcsec from the targets. {\tt ForcePho} fits these profiles simultaneously and with full posterior sampling, allowing us to measure the uncertainties in the profile and the covariance of the fluxes between sources that appear blended, such as JADES-GS-z14-0 and its low redshift neighbor. 

\begin{figure}
    \centering
        \includegraphics[width=0.9\linewidth,
    trim=0cm 0cm 0cm 0cm, clip=true]{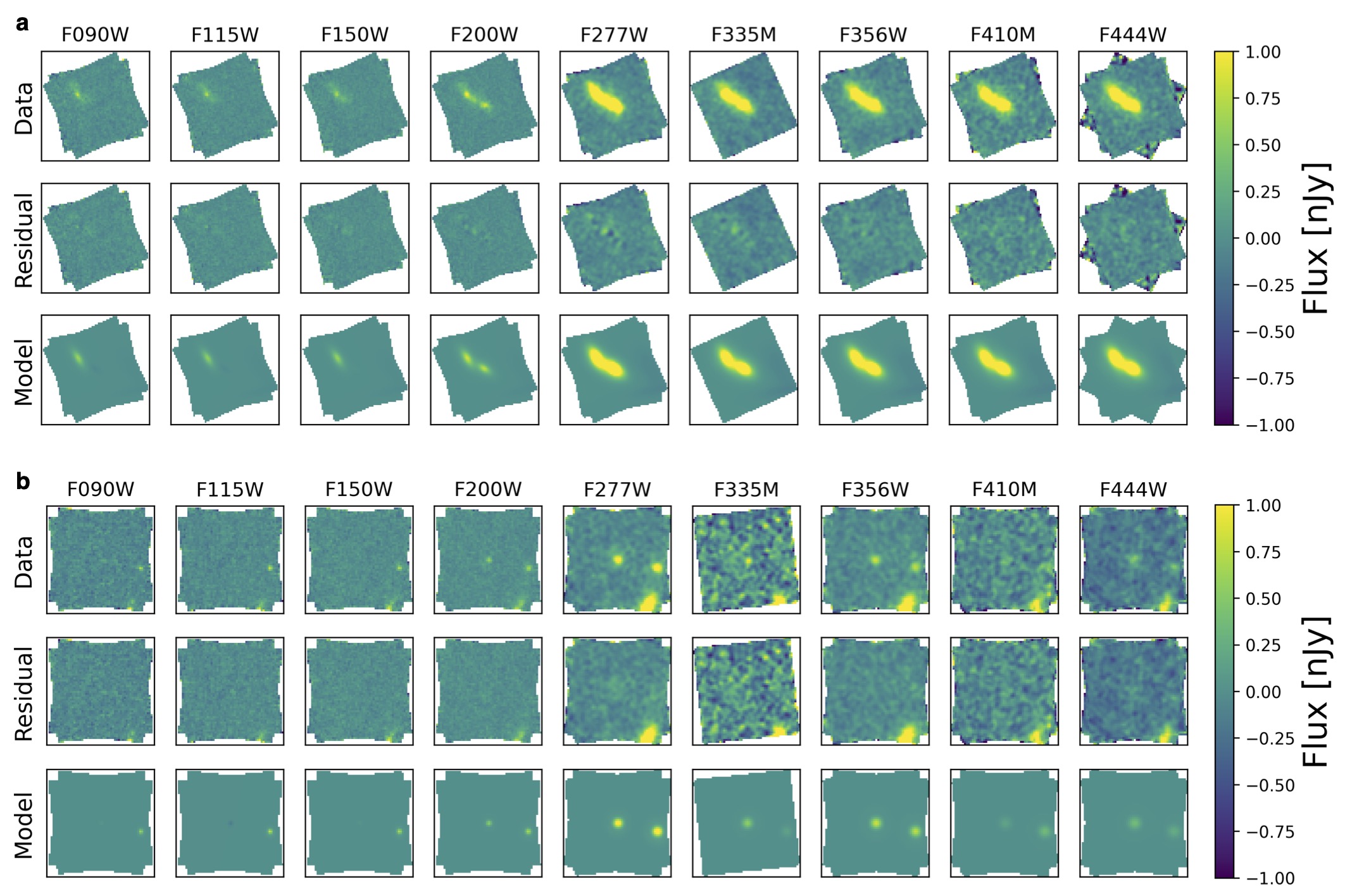}
    \caption{{\bf ForcePho model fitting. a--b,} The data, residual, model, and fluxes with the recovered galaxy component for JADES-GS-z14-0 ({\bf a} ) and JADES-GS-z14-1 ({\bf b}) in the multiple NIRCam bands. Each frame shows a $2^{\prime\prime}\times 2^{\prime\prime}$ stamp centred on the location of each galaxy. The figure shows that the model has fit the data well within all bands without leaving significant residuals.}
    \label{fig:forcepho}
\end{figure}

Data, residual, and model are in Extended Data Fig.~\ref{fig:forcepho}. A single component is sufficient to match the data of both JADES-GS-z14-0 and JADES-GS-z14-1.
In both cases, the surface brightness profile is consistent with an S\'{e}rsic profile with an index of $\sim1$. For the brightest of the two galaxies, we have determined a de-convolved half-light radius of 260 pc, while the compact size of JADES-GS-z14-1 has returned just an upper limit of 160 pc.

\subsubsection*{Dust enrichment}

\begin{figure}
    \centering
    \includegraphics[width=0.5\linewidth]{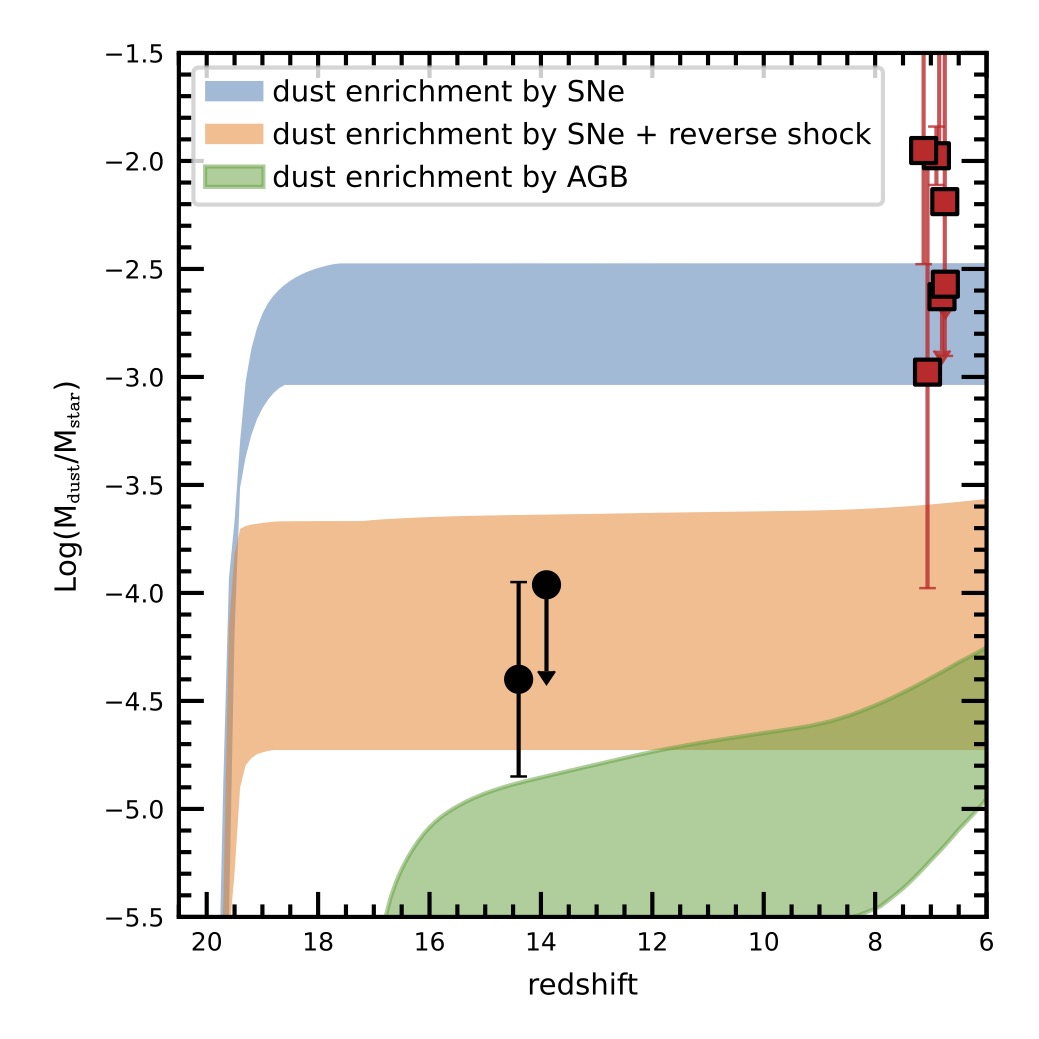}
    
    \caption{{\bf Dust to stellar mass ratio.} Dust enrichment models by supernovae (SNe) with no (blue) and with (orange) the effect of the partial dust destruction due to the reverse shock, and asymptotic giant branch (AGB) stars from a single star-formation burst at $z=20$. The relative contribution of the SNe and AGB dust formation are estimated assuming an initial progenitor metallicity in the range $0.01 < Z/Z_\odot < 0.1$  \cite{Schneider:2023}. Black circles illustrate the dust-to-stellar mass ratio of JADES-GS-z14-0 and JADES-GS-z14-1. A proxy of the dust masses of the two $z\sim14$ galaxies has been estimated from the $A_\mathrm{ V}$ parameter, assuming that dust is distributed in a sphere with a radius as large as the galaxy radius \cite{Ferrara:2023}. We also report in red squares the dust-to-stellar mass ratio of HST-selected, UV-bright galaxies at $z\sim6\text{--}7$, with dust measurements from ground-based millimeter telescopes \cite{Witstok:2023, Witstok:2024, Valentino:2024}.}

    \label{fig:dust}
\end{figure}

Following Ferrara et al.\cite{Ferrara:2023} and assuming the prescription Calzetti et al.\cite{Calzetti:2000} for the dust attenuation law, $A_{V}$ could be used as a proxy for the dust mass: $M_d \approx  2.2\times 10^4 A_{V}({r_\mathrm{ d}/\mathrm{ 100 pc}})^2~\mathrm{ M_\odot}$, where ${r_d}$ is the radius over which the interstellar medium dust is assumed to extend. However, we stress that the $A_{V}$ parameter is not an accurate measure of the actual amount of dust as this parameter is simply estimated from the spectral fitting assuming an dust attenuation law.
Assuming that the spatial extension of the dust is as large as the size of the galaxy (i.e. $r_d=r_\mathrm{ UV}$), we derived a dust mass $M_d= 5\times 10^4~\mathrm{ M_\odot}$ and $M_d< 0.5\times 10^4~\mathrm{ M_\odot}$ for the two galaxies, respectively. Comparing these masses with the stellar masses, we find a dust-to-stellar mass fraction of $<10^{-4}$ for both galaxies (Extended Data Fig.~\ref{fig:dust}), which is a factor $>10$ lower than those inferred in galaxies at $z\sim6-7$ \cite[$10^{-2}-10^{-3}$][]{Witstok:2023, Witstok:2023a, Valentino:2024} and that predicted by supernova (SN) models without reverse shock \cite{Todini:2001, Schneider:2023}. Extended Data Figure~\ref{fig:dust} shows the timescales to reach the asymptotic dust-to-stellar mass ratio due to various dust enrichment processes and assuming that galaxies formed at $z=20$ from a single star-formation burst (see review by Schneider et al.\cite{Schneider:2023}). In the first few Myrs, the dust is dominated by SNe, and in less than 5 Myr, the galaxy has already reached a dust-to-stellar mass ratio of $\sim1/1000$. Even assuming a different realistic star-formation history, the dust-to-stellar mass reaches the asymptotic values of $<10$~Myr. Reverse shocks created by the interaction between the expanding SN blast wave and the interstellar medium can limit the effective dust enrichment by SNe\cite{Bocchio:2016, Marassi:2019, Graziani:2020, DiCesare:2023}, as indicated by the orange shaded region in Extended Data Figure~\ref{fig:dust}. However, the efficiency of dust destruction due to reverse shocks is still debated, and different models predict quite different survival rates \cite{Schneider:2023}.

\subsubsection*{Acknowledgments}
 SC, EP \& GV acknowledge support by European Union’s HE ERC Starting Grant No. 101040227 - WINGS.
SA acknowledges support from the JWST Mid-Infrared Instrument (MIRI) Science Team Lead, grant 80NSSC18K0555, from NASA Goddard Space Flight Center to the University of Arizona.
RM, WB, FDE, JW \& JS acknowledge support by the Science and Technology Facilities Council (STFC), ERC Advanced Grant 695671 ``QUENCH", and by the UKRI Frontier Research grant RISEandFALL. RM also acknowledges funding from a research professorship from the Royal Society.
This research is supported in part by the Australian Research Council Centre of Excellence for All Sky Astrophysics in 3 Dimensions (ASTRO 3D), through project number CE170100013.
AJB, AJC, JC, AS \& GCJ acknowledge funding from the ``FirstGalaxies" Advanced Grant from the European Research Council (ERC) under the European Union’s Horizon 2020 research and innovation programme (Grant agreement No. 789056).
ECL acknowledges support of an STFC Webb Fellowship (ST/W001438/1).
DJE, EE, BDJ, GR, MR, FS, and CNAW are supported by JWST/NIRCam contract to the University of Arizona NAS5-02015.
DJE is also supported as a Simons Investigator.
The Cosmic Dawn Center (DAWN) is funded by the Danish National Research Foundation under grant DNRF140.
BER acknowledges support from the NIRCam Science Team contract to the University of Arizona, NAS5-02015, and JWST Program 3215.
B.R.P. acknowledges support from the research project PID2021-127718NB-I00 of the Spanish Ministry of Science and Innovation/State Agency of Research (MICIN/AEI/ 10.13039/501100011033).
RS acknowledges support from a STFC Ernest Rutherford Fellowship (ST/S004831/1).
ST acknowledges support by the Royal Society Research Grant G125142.
H{\"U} gratefully acknowledges support by the Isaac Newton Trust and by the Kavli Foundation through a Newton-Kavli Junior Fellowship.
The research of CCW is supported by NOIRLab, which is managed by the Association of Universities for Research in Astronomy (AURA) under a cooperative agreement with the National Science Foundation.
PGP-G acknowledges support from grant PID2022-139567NB-I00 funded by Spanish Ministerio de Ciencia e Innovaci\'on MCIN/AEI/10.13039/501100011033, FEDER, UE.

\subsubsection*{Author contributions}
SC, FDE, and PJ contributed to the analysis, and initial interpretation of the spectroscopic data. 
All authors contributed to the interpretation of results. SC, SA, PJ, MC, JW, EP, and GV contributed to the NIRSpec data reduction and to the development of the NIRSpec pipeline. PJ, CW, AB, and KH contributed to the design and optimization of the MSA configurations. DJE, BDJ, BR, CW, and ST contributed to the analysis and interpretation of the NIRCam imaging data. 
DJE, JMH, and GR contributed to the analysis and interpretation of the MIRI imaging data.
PJ, JW, and FDE. contributed to the development of tools for the spectroscopic data analysis. 

\subsubsection*{Data availability}
The NIRCam data that support the findings of this study are publicly available at https://archive.stsci.edu/hlsp/jades. The reduced spectra that support the findings of this study are publicly available on https://doi.org/10.5281/zenodo.12578543

\subsubsection*{Code availability}
The AstroPy software suite is publicly available, as ForcePho. BEAGLE is available via a Docker image upon request at http://www.iap.fr/beagle/.

\clearpage

\newcommand{\actaa}{Acta Astron.}   
\newcommand{\araa}{Annu. Rev. Astron. Astrophys.}   
\newcommand{\areps}{Annu. Rev. Earth Planet. Sci.} 
\newcommand{\aar}{Astron. Astrophys. Rev.} 
\newcommand{\ab}{Astrobiology}    
\newcommand{\aj}{Astron. J.}   
\newcommand{\ac}{Astron. Comput.} 
\newcommand{\apart}{Astropart. Phys.} 
\newcommand{\apj}{Astrophys. J.}   
\newcommand{\apjl}{Astrophys. J. Lett.}   
\newcommand{\apjs}{Astrophys. J. Suppl. Ser.}   
\newcommand{\ao}{Appl. Opt.}   
\newcommand{\apss}{Astrophys. Space Sci.}   
\newcommand{\aap}{Astron. Astrophys.}   
\newcommand{\aapr}{Astron. Astrophys. Rev.}   
\newcommand{\aaps}{Astron. Astrophys. Suppl.}   
\newcommand{\baas}{Bull. Am. Astron. Soc.}   
\newcommand{\caa}{Chin. Astron. Astrophys.}   
\newcommand{\cjaa}{Chin. J. Astron. Astrophys.}   
\newcommand{\cqg}{Class. Quantum Gravity}    
\newcommand{\epsl}{Earth Planet. Sci. Lett.}    
\newcommand{\expa}{Exp. Astron.}    
\newcommand{\frass}{Front. Astron. Space Sci.}    
\newcommand{\gal}{Galaxies}    
\newcommand{\gca}{Geochim. Cosmochim. Acta}   
\newcommand{\grl}{Geophys. Res. Lett.}   
\newcommand{\icarus}{Icarus}   
\newcommand{\ija}{Int. J. Astrobiol.} 
\newcommand{\jatis}{J. Astron. Telesc. Instrum. Syst.}  
\newcommand{\jcap}{J. Cosmol. Astropart. Phys.}   
\newcommand{\jgr}{J. Geophys. Res.}   
\newcommand{\jgrp}{J. Geophys. Res.: Planets}    
\newcommand{\jqsrt}{J. Quant. Spectrosc. Radiat. Transf.} 
\newcommand{\lrca}{Living Rev. Comput. Astrophys.}    
\newcommand{\lrr}{Living Rev. Relativ.}    
\newcommand{\lrsp}{Living Rev. Sol. Phys.}    
\newcommand{\memsai}{Mem. Soc. Astron. Italiana}   
\newcommand{\maps}{Meteorit. Planet. Sci.} 
\newcommand{\mnras}{Mon. Not. R. Astron. Soc.}   
\newcommand{\nat}{Nature} 
\newcommand{\nastro}{Nat. Astron.} 
\newcommand{\ncomms}{Nat. Commun.} 
\newcommand{\ngeo}{Nat. Geosci.} 
\newcommand{\nphys}{Nat. Phys.} 
\newcommand{\na}{New Astron.}   
\newcommand{\nar}{New Astron. Rev.}   
\newcommand{\physrep}{Phys. Rep.}   
\newcommand{\pra}{Phys. Rev. A}   
\newcommand{\prb}{Phys. Rev. B}   
\newcommand{\prc}{Phys. Rev. C}   
\newcommand{\prd}{Phys. Rev. D}   
\newcommand{\pre}{Phys. Rev. E}   
\newcommand{\prl}{Phys. Rev. Lett.}   
\newcommand{\psj}{Planet. Sci. J.}   
\newcommand{\planss}{Planet. Space Sci.}   
\newcommand{\pnas}{Proc. Natl Acad. Sci. USA}   
\newcommand{\procspie}{Proc. SPIE}   
\newcommand{\pasa}{Publ. Astron. Soc. Aust.}   
\newcommand{\pasj}{Publ. Astron. Soc. Jpn}   
\newcommand{\pasp}{Publ. Astron. Soc. Pac.}   
\newcommand{\raa}{Res. Astron. Astrophys.} 
\newcommand{\rmxaa}{Rev. Mexicana Astron. Astrofis.}   
\newcommand{\sci}{Science} 
\newcommand{\sciadv}{Sci. Adv.} 
\newcommand{\solphys}{Sol. Phys.}   
\newcommand{\sovast}{Soviet Astron.}   
\newcommand{\ssr}{Space Sci. Rev.}   
\newcommand{\uni}{Universe} 

\bibliography{z14_biblio}

\end{document}